\documentclass[]{raa}                  % preprint format
\usepackage{graphicx,times}             %for PS/EPS graphics inclusion, new
\input{epsf.sty}                        %for PS/EPS graphics inclusion, old
\input{psfig.sty}                       %for PS/EPS graphics inclusion, old

\usepackage{graphicx}
\usepackage{lscape}
\usepackage{longtable}
\usepackage[usenames]{color}

\def\alpox{$\alpha_{\rm ox}$ }

\def\alprx{$\alpha_{\rm rx}$ }
\def\alprxe{$\alpha_{\rm rx}$}
\def\alpro{$\alpha_{\rm ro}$ }
\def\alproe{$\alpha_{\rm ro}$}

\def\alpxe{$\alpha_{\rm x}$}

\def\alpoe{$\alpha_{\rm o}$}

\def\hb{H$\beta$ }
\def\hbe{H$\beta$}
\def\ha{H$\alpha$ }
\def\hae{H$\alpha$}

\def\lr{$L_{\rm r}$}
\def\lrc{$L_{\rm 5GHz}^{\rm core}$}
\def\lblr{$L_{\rm BLR}$}

\def\mbh{$M_{\rm BH}$}

\def\msune{\ensuremath{M_{\odot}}}

\def\ulume{${\rm erg\,s^{-1}}$}

\def\kmps{${\rm km\,s^{-1}}$ }
\def\kmpse{${\rm km\,s^{-1}}$}
\def\whz{$\rm W\,Hz^{-1}$ }

\begin{document}

   \title{Optical identification of radio-loud active galactic nuclei in the ROSAT-Green-Bank sample with SDSS spectroscopy}
%\footnotetext{$*$ Supported by the National Natural Science Foundation of China.}

%   \subtitle{I. Place Your Subtitle Here}

   \volnopage{Vol.0 (200x) No.0, 000--000}      %%preserved for Editor. DOn't remove!
   \setcounter{page}{1}          %%starting page, preserved for Editor. DOn't remove!

   \author{De-Liang Wang
      \inst{1,2}
   \and Jian-Guo Wang
      \inst{1,2}
   \and Xiao-Bo Dong
      \inst{3}
   }

   \institute{
             National Astronomical Observatories/Yunnan Observatory,
             Chinese Academy of Sciences, Kunming, Yunnan, 650011, China;
             {\it dlwang@mail.ynao.ac.cn}\\
        \and
%             Full institute address for the second author\\
             Graduate School of the Chinese Academy of Sciences, 19A Yuquan Road, Beijing 100039, China
        \and
%             Full institute address for the third author\\
             Center for Astrophysics, University of Science and Technology of China, Hefei, 230026, China
   }

   \date{Received~~2009 month day; accepted~~2009~~month day}

\abstract{ Results of extended and refined optical identification of
181 radio/X-ray sources in the RASS-Green Bank (RGB) catalog
 (Brinkmann et~al.~\cite{Brinkmann97}) are
presented which have been spectroscopically observed in the Sloan
Digital Sky Survey (SDSS) DR5. The SDSS spectra of the optical
counterparts are modeled in a careful and self-consistent way by
incorporating the host galaxy's starlight. The optical emission line
parameters are presented, which are derived accurately and reliably,
along with the radio 1.4--5\,GHz spectral indices estimated using
(non-simultaneous) archival data. For 72 sources, the
identifications are presented for the first time. It is confirmed
that the majority of strong radio/X-ray emitters are radio-loud
active galactic nuclei (AGNs), particularly blazars. Taking
advantage of the high spectral quality and resolution and our
refined spectral modeling, we are able to disentangle narrow line
radio galaxies (NLRGs), as vaguely termed in most previous
identification work, into Seyfert\,II galaxies and LINERs
(low-ionization nuclear emission regions), based on the standard
emission line diagnostics. The NLRGs in the RGB sample, mostly
belonging to `weak line radio galaxies' (Tadhunter et
al.~\cite{tad98}), are found have optical spectra consistent
predominantly with LINERs, and only a small fraction with
Seyfert\,II galaxies. A small number of LINERs have radio power as
high as $10^{23}-10^{26}$\,\whz at 5\,GHz, being among the strongest
radio emitting LINERs known so far. Two sources are identified with
radio-loud narrow line Seyfert 1 galaxies (NLS1s), a class of rare
objects. The presence is also confirmed of flat-spectrum radio
quasars whose radio--optical--X-ray effective spectral indices are
similar to those of High-energy peaked BL Lacs (HBLs), as suggested
by Padovani et~al.~(\cite{P03}), albeit it is yet a debate as to
whether this is the case for their actual spectral energy
distributions.  }
   \authorrunning{De-Liang Wang, Jian-Guo Wang, Xiao-Bo Dong }            %author_head in even pages
   \titlerunning{Optical identification of radio-loud AGNs in the RGB sample}  % title_head in odd pages

   \maketitle
%% The author head (on even pages) and the title head (on odd pages) will be
%% automatically extracted from \author{} and \title{}. Whenever the title is too long,
%% you will be asked to supply a shorter one by inserting either \authorrunning{} or
%% \titlerunning{} before \maketitle. Anyway, you can specify your own heads.
%%
%%
%% Note: In the following text body of your manuscript, please note several differences from
%%       other major journals:
%% (1) \subsection{Please Capitalize the First Letter of Each Notional Word in Subsection Title}
%% (2) Please Capitalize the First Letter of Each Notional Word in all tables' captions

%
%________________________________________________ sections below
%
\section{Introduction}           %% first-level sections will be auto-capitalized
\label{sect:intro}

Sources detected in both X-ray and radio surveys are predominantly
radio emitting active galactic nuclei (AGNs), especially blazars.
Optical identification of such sources has proved to be an efficient
way to compile radio-loud AGNs and blazar samples. With the advent
of the ROSAT satellite---both the ROSAT all-sky survey (RASS, Voges
et~al.~\cite{vog}) and pointed observations---in the X-ray band and
several large area sky surveys in the radio bands, large samples of
radio-loud AGNs and blazars have been discovered and identified.
Examples of such samples are the RGB (RASS--Green Bank) sample
comprising 2127 sources (Brinkmann et~al.~\cite{Brinkmann95},
~\cite{Brinkmann97}; Laurent-Muehleisen et~al.~\cite{LM98},
\cite{LM99}), the REX (radio-emitting X-ray) sample (Caccianiga
et~al.~\cite{Caccianiga99}, \cite{Caccianiga2000}), the DXRBS (Deep
X-ray radio blazar survey) sample (Perlman et~al.~\cite{Per98},
Landt et~al.~\cite{L01}), the ``Sedentary'' sample (Giommi
et~al.~\cite{Gio99a}, \cite{Gio05}; Piranomonte
et~al.~\cite{Pir07}), and the ROXA sample (Turriziani
et~al.~\cite{Tur07}), etc., which were compiled  using various
selection criteria. Objects identified optically in these samples
include radio quasars, BL Lac objects, broad-line and narrow-line
radio galaxies, and even clusters of galaxies. For most of these
samples, the optical identification is far from complete; more
identifications still await new optical spectroscopic observations.

In most previous identification studies of large samples of
radio/X-ray sources, the spectral classification was generally
rather coarse, given generally low spectral resolution and/or
signal-to-noise ratio (S/N) acquired, and/or poor spectrophotometry
calibration. For instance, weak or relatively narrow broad-line
components were hard to detect, leading to incorrect classifications
between type\,I and type\,II AGNs. Moreover, emission line objects
without an apparent broad line component were often classified
collectively as narrow line radio galaxies (NLRGs), which are
commonly thought to be the radio-loud counterparts of Seyfert\,II
galaxies (e.g. Osterbrock~\cite{ost76}; Koski~\cite{kos78}; Urry \&
Padovani~\cite{urry95}). However, this practice is vague and
sometimes misleading, since it has been known for a long time that
some radio galaxies have optical spectra that hardly qualify as
NLRGs (or Seyfert\,II galaxies). Even from the early work on
identification of radio sources, some radio galaxies were found to
show spectra dominated by low-ionization lines, unlike Seyfert\,II
galaxies (e.g. Costero \& Osterbrock~\cite{cos77}; Laing et
al.~\cite{lai94}), especially in FR\,I sources (e.g. Hine \&
Longair~\cite{hin79}; Laing et al.~\cite{lai94}). On the other hand,
there also exists a distinct group of radio sources with weak [O
III] line strength, termed as weak line radio galaxies (WLRG,
Tadhunter et al.~\cite{tad98}). In fact, most of these objects, when
observed with a better spectral resolution and S/N, turned out to be
LINERs (Lewis et al.~\cite{lew03}). Rigorous classification of
emission line galaxies should rely on the emission line ratio
diagnostics (e.g. Baldwin et~al.~\cite{bpt}; Veilleux \&
Osterbrock~\cite{vo87}). This, however, requires good
spectrophotometric calibration and sufficient spectral resolution
and S/N, which could not be reached in some of the identification
work of  radio and X-ray sources in the past.

Furthermore, the ground-based slit or fiber spectra of moderately or
under- luminous AGNs are often contaminated by starlight of the host
galaxies. The spectra of host galaxy starlight could severely affect
the detection and measurement of emission lines, by suppressing some
weak emission lines (such as $H\beta$), distorting the line fluxes,
and even generating spurious weak broad emission lines.

The Sloan Digital Sky Survey (SDSS, Stoughton et~al.~\cite{st02})
has acquired optical spectra of about one million galaxies and
quasars with homogeneous quality and reasonably good spectral
resolution, as well as good spectrophotometric calibration accuracy
($\sim 8\%$, Vanden~Berk et~al.~\cite{Van04}). Furthermore,
efficient algorithms have been developed to decompose nuclear
emission line spectra from the host galaxy starlight, making it
possible to accurately measure emission line fluxes and to recover
weak emission line features. These improvements allow refined
optical spectral classification of radio or X-ray sources. This is
particularly relevant to low-luminosity AGNs (LLAGNs) with
relatively weak emission lines. Identification of these sources with
LLAGNs would complement the study of LLAGNs in general by providing
useful radio and X-ray data.

In recent studies of X-ray--radio AGN samples, a few interesting new
trends were found. One was the suggestion of the presence of
flat-spectrum radio quasars (FSRQs) appearing to show spectral
energy distributions (SEDs) typical of high-energy peaked BL Lac
objects (HBLs) (e.g. Padovani et~al.~\cite{P03}; Perlman
et~al.~\cite{Per98}). Such objects were claimed to be mostly found
in X-ray selected samples for their strong X-ray emission, and were
missed in earlier pure radio-selected blazar samples. If the SEDs of
these objects indeed similar to those of HBLs, they would present a
challenge to the so-called ``blazar sequence''---a model in which
the blazar SED is strongly correlated with the source power (Fossati
et~al.~\cite{Fos98}; Ghisellini et al.~\cite{ghi98}). This is
because objects with HBL-type SED at quasar luminosities are not a
consequence of the model (Padovani et~al.~\cite{P03}). Independent
confirmation of the presence of such FSRQs would still be
worthwhile, although it is still a subject to actively debate
whether the  effective two-point spectral indices are good
indicators of HBL-type SEDs (Maraschi et al.~\cite{mar08}).
Moreover, recently Yuan et al.~(\cite{yuan08}) found a fraction of
very radio-loud narrow line Seyfert 1 galaxies (NLS1) in their small
sample appearing to have  HBL-type effective spectral indices
(though at relatively low luminosities). A related question then is
what is the fraction of NLS1 in FSRQs with ``apparent'' HBL-type
SEDs. This can only be addressed by observations with sufficient S/N
ratios and spectral resolutions that resolve a relatively narrow
broad-component ($<2000$\,\kmpse) in the presence of (strong) narrow
Balmer lines.

Motivated by the above issues, we carry out optical identification
of unidentified sources in the RGB sample (Brinkmann
et~al.~\cite{Brinkmann97}) using the SDSS spectroscopic data. The
choice of the RGB sample is based on the following considerations.
Firstly, the sample has refined VLA astrometry accuracy of $\pm
0.5$\,$''$ to ensure unambiguous matching with SDSS objects; also
there is a large overlap of sky coverage of the RGB sample with the
SDSS spectroscopic survey. Secondly, the RGB's high X-ray flux limit
preferentially selects X-ray bright objects. This is essential for
selecting FSRQs with ``apparent'' HBL-type spectral indices (Perlman
et~al.~\cite{Per98}; Padovani et~al.~\cite{P03}). Also, given its
high thresholds in both the radio and X-ray bands, any
identification with LLAGNs would place the objects in the regime at
the higher-end of the luminosity function of LLAGNs, which are
otherwise difficult to find owing to their rarity. We describe the
RGB--SDSS sample and optical data analysis in Section\,2, and
present the results  in Section\,3. The results are summarized and
discussed in Section\,4, with  emphases given to LLAGNs and FSRQs
with HBL-type spectral indices. Through out the paper, we use a
cosmology with $H_{0}$=70 km $s^{-1}$ ${Mpc}^{-1}$,
${\Omega}_{M}$=0.3, and ${\Omega}_{\Lambda}$=0.7.

\section{RGB-SDSS SAMPLE and optical spectral analysis}
\subsection{RGB-SDSS sample}
The RGB catalog (Brinkmann et~al.~\cite{Brinkmann95})
was constructed by cross-correlating the ROSAT All-Sky Survey
 (RASS) with the radio catalog created from the 1987 Green Bank (GB) survey
maps (Gregory \& Condon~\cite{Gre91}, Gregory et~al.~\cite{Gre96}).
The radio catalog consists of sources of $\geq$ 3$\sigma$ confidence
with a flux limit ranging from 15 mJy  to 24 mJy (at low
declinations). The initial RGB sample is composed of  2127 matches
with an angular separation $<$100$''$ between the X-ray and radio
positions. Among them, 617 sources had been previously optically
identified as extragalactic objects using NED by Brinkmann
et~al.~(\cite{Brinkmann95}). The sample sources were further
observed with the VLA to pin down their radio positions to an
accuracy of $\pm 0.5''$ (Laurent-Muehleisen et~al.~\cite{LM97}).
Using the accurate VLA radio positions, a refined large sample of
1304 X-ray/radio sources from the RGB sample, mostly optically
unidentified \footnote{In another identification attempt,
Laurent-Muehleisen et~al.~(\cite{LM98}) spectroscopically observed
and identified 169 objects with the McDonald 2.7m and the Kitt Peak
2.1m telescopes.}, was presented in Brinkmann
et~al.~(\cite{Brinkmann97}). This sample forms the parent sample of
this study. We cross-correlated all the radio sources associated
with the 1304 RGB X-ray sources in Brinkmann
et~al.~(\cite{Brinkmann97}) with the SDSS DR5 spectroscopic
databases of both galaxies and quasars. A matching criterion of
3$''$ was used. This yielded 181 objects with SDSS spectra, 113 of
which were assigned as QSOs, 49 as galaxies and 19 as unknown {\em
by the SDSS pipeline}.

To estimate the radio spectral indices\footnote{We assume a power
law form $f_{\nu} \sim {\nu}^{-\alpha_r}$. } of the objects, we also
searched for their emission at 1.4\,GHz from the NVSS (NRAO VLA Sky
Survey). A matching criterion of 3$'$ \,in radius was used when
cross-correlating the VLA and NVSS positions. Since the GB 5\,GHz
survey had a resolution (beam size, 3.6$'$$\times$3.4$'$) larger
than that of NVSS, for each match, we consider the following cases
respectively to eliminate inconsistency in the estimation of source
fluxes caused by the effect of different angular resolutions. First,
if the NVSS counterpart is the only radio source within 3$'$ of the
VLA position, the NVSS data is simply considered as its 1.4\,GHz
flux. This is the case for the vast majority of sample objects (86).
Second, we consider the case where there are more NVSS sources than
matching NVSS counterparts at the VLA positions within 3$'$. If the
summed flux of the former is significantly weaker than the latter,
we used the total 1.4\,GHz radio flux within 3$'$ to calculate the
spectral index; otherwise, we consider that the object's 5\,GHz flux
is significantly contaminated by nearby sources and subsequently no
spectral index can be inferred. As a result, we obtained the radio
indices for 103 objects. Following the convention, flat- and
steep-spectrum sources have ${\alpha}_r \leq $ 0.5 and ${\alpha}_r >
$ 0.5, respectively. The sample is summarized in Table\,1, with some
of the optical and radio data. The redshifts are measured from our
own spectral data analysis (see below).

\subsection{Optical spectral analysis}
SDSS spectra were
taken through fibers with an aperture of 3$''$ in diameter and
 have a wavelength coverage from 3800 to 9200\AA\,
and a resolution of R$\sim$1800--2200. For some objects, especially
those assigned as galaxies by the SDSS pipe-line, their spectra
include significant amounts of (or even are dominated by) starlight
from the host galaxies. Thus, the removal of the stellar spectrum is
essential for the proper detection and measurement of nuclear
emission lines, especially when the lines, broad or narrow, are not
strong. This is important for ensuring reliable spectral
classification of identified objects. We consider objects as having
non-negligible star-light contribution if their spectra show
(possible) absorption lines of Ca\,K (3934\AA), Ca\,K + H$\epsilon$
(3970\AA), or the high-order Balmer line H$\delta$ (with detection
significance $> 2\sigma$). For proper modeling of host galaxy
starlight, we used the algorithm developed at the University of
Science of Technology of China, which was described in detail in
Zhou et~al.~(\cite{zhou06}) (see also Lu et~al.~\cite{lu06}) and is
summarized in Appendix\,A. The modeled host galaxy stellar spectrum
is then subtracted from the SDSS spectrum.

For most of the RGB-SDSS objects, the SDSS spectra
or the leftover spectra (modeled starlight spectrum subtracted)
 show significant  emission lines.
The emission line spectra are fitted using an updated version of the
code as described in Dong et~al.~(\cite{dong05}), with improvements
made for optimized recovery of weak emission lines. The method of
the spectral fitting is also summarized in Appendix\,A. The Balmer
emission lines are de-blended into a narrow and a broad component if
the latter is present. If a broad line component is detected at the
$\geq 5 \sigma$ confidence level, we regard it as genuine. As shown
in Zhou et~al.~(\cite{zhou06}), this analysis procedure yields
reliable detection of emission lines and accurate measurements of
the line parameters. The derived parameters of the common emission
lines are given in Table\,\ref{tab:nlelp} and Table\,\ref{bel} for
the narrow- and broad-line components, respectively.

\section{Results of optical identification and classification}

Thanks to the accurate astrometry of radio VLA positions and optical
SDSS positions, the optical identifications of the RGB-SDSS objects
are unambiguous. Here, we present detailed spectral classification
of the 181 RGB-SDSS objects (see table\,\ref{spec_list}). The sample
is first separated into two groups: (i) emission-line objects with
at least a few common significantly detected emission lines (e.g.
$H\alpha$, $H\beta$, [O III], [N II], [S II]) detected
significantly, with signal-to-noise ratio S/N $>3$ and equivalent
width $EW \ge$ 5\AA\ for at least one emission line (127 sources);
 and (ii) objects with weak or no emission lines (54 sources)
(see below for the $EW$ threshold distinguishing Seyfert galaxies
and BL Lac objects).

\begin{table}
\begin{center}
\caption{Source number of optical identification and classifications
for the RGB-SDSS sample \label{spec_list}}
\begin{tabular}{l|r}
\hline
classification &number\\
(1) & (2)\\
\hline
QSO(type I)&84\\
Seyfert I&27\\
NLSy1    &2\\
LINERs   &12\\
Seyfert II&2\\
BL Lac   &32\\
BL Lac candidate&11\\
Galaxy   &11\\
\hline
\end{tabular}
\end{center}
\end{table}

\subsection{Broad line AGNs}
\begin{figure}
\begin{center}
 \includegraphics[width=1.0\hsize,height=0.4\hsize]{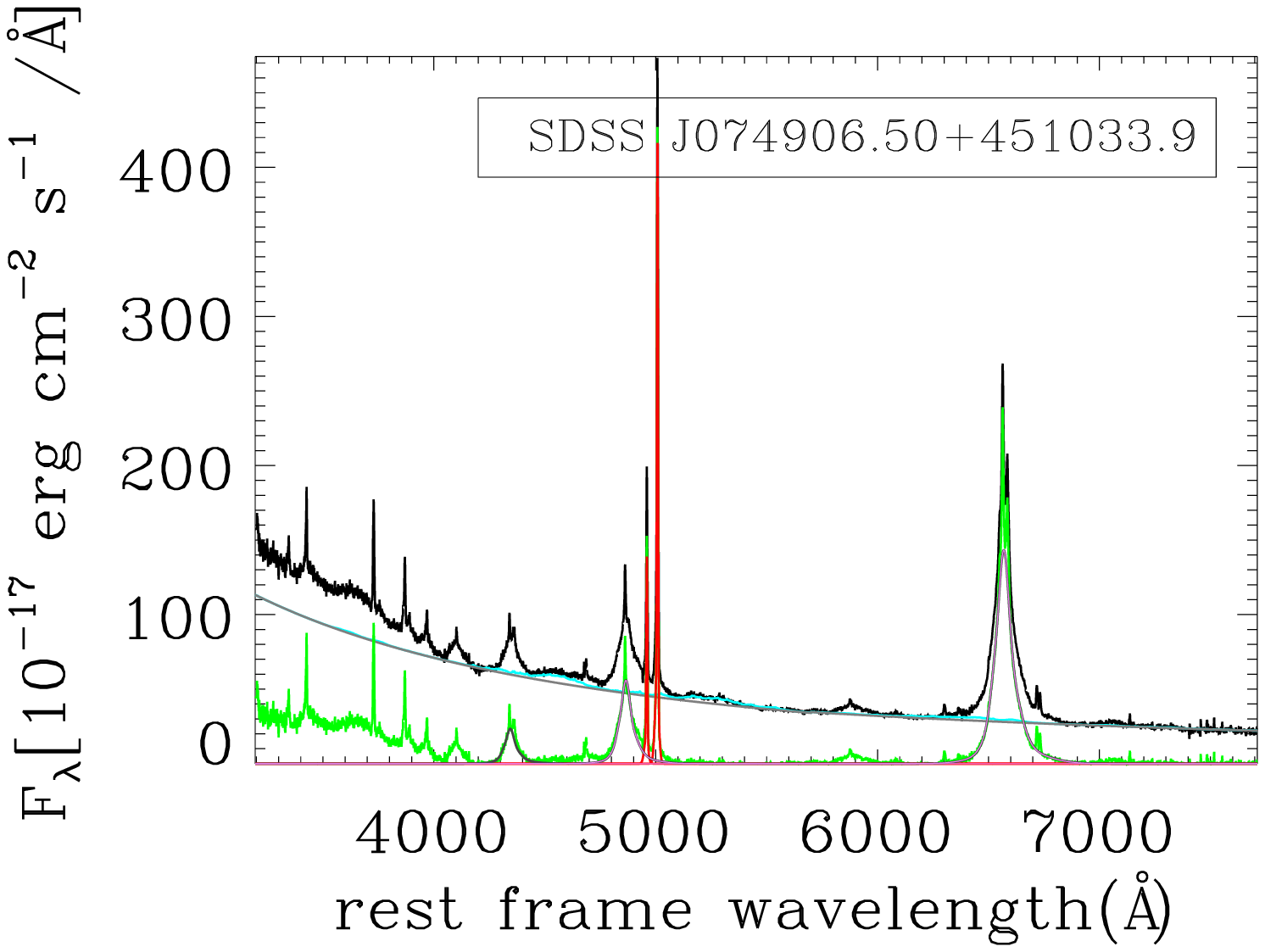}
 \includegraphics[width=1.0\hsize,height=0.4\hsize]{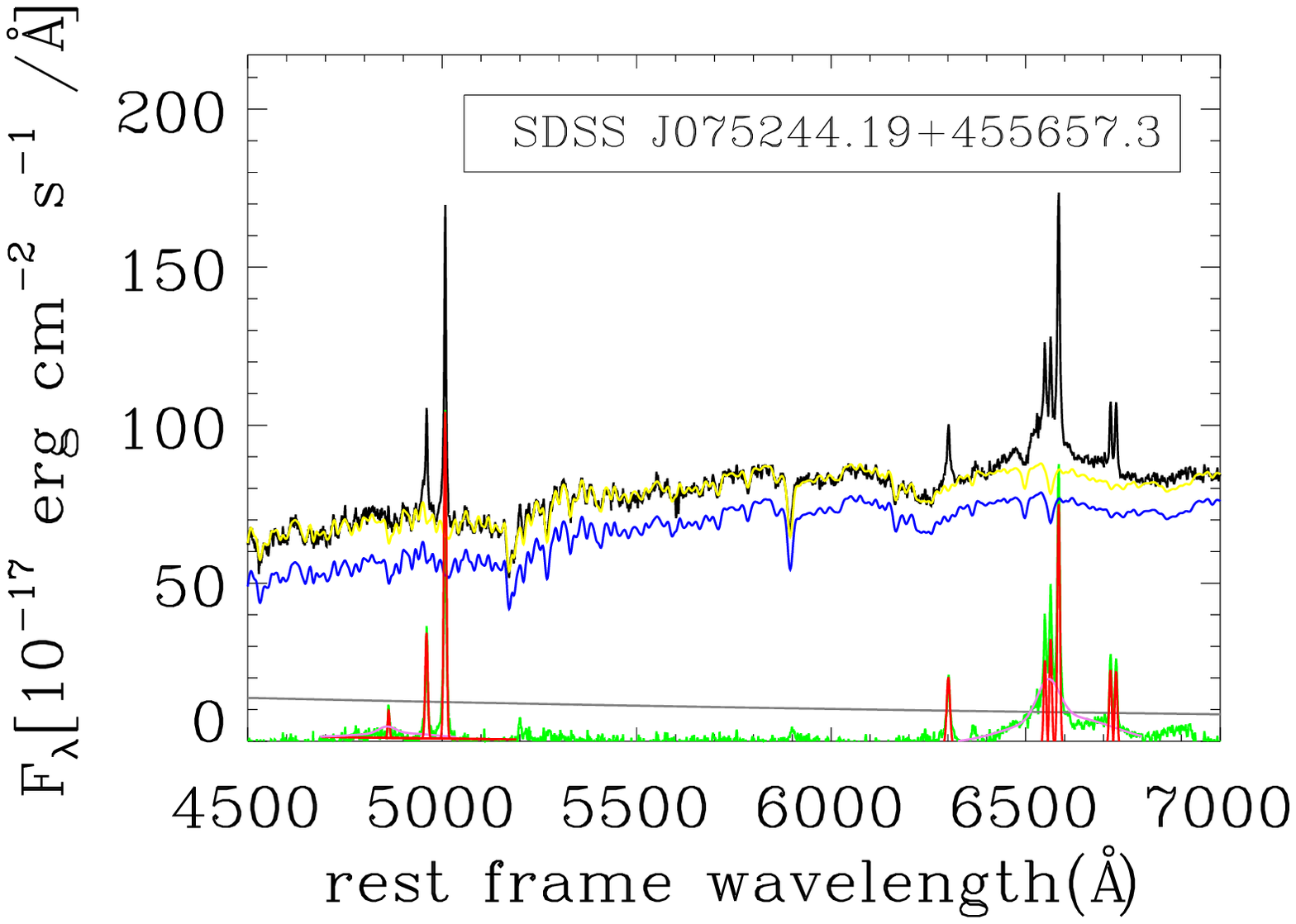}
 \caption{\label{fig:spec_blagn}
Examples of the rest frame spectra (black) of identified broad line
(type\,I) AGNs for a quasar (upper panel) and a BLRG (lower panel,
with nuclear--starlight decomposition performed). Also plotted are
the best-fit model components: starlight (blue), nuclear power-law
continuum (grey) with the Fe\,II multiplets superimposed (cyan), the
nuclear continuum plus starlight (yellow), residuals after
subtraction of the nuclear continuum and starlight (green), broad
(purple) and narrow (red) emission line components. }
\end{center}
\end{figure}

Of the 113 emission line objects with a significant broad component
(see table \ref{bel}), there are 91 that have a broad component of
the Balmer lines detected at S/N$>5$, and the others with Mg II
broad emission lines. By broad component we mean that its line width
is significantly broader than that of the corresponding narrow
component of the line, typically several hundred kilo-meters per
second. The smallest width of the broad component found in our
sample is 1500\,\kmps, still much broader than the narrow lines. We
collectively classify these objects as broad line AGNs (BLAGNs),
i.e.\ or type\,I AGN. They are further separated into quasars and
Seyfert\,I galaxies, using the common absolute magnitude dividing
line\footnote{We used $M_{\rm B}=-21.5+5log (H_0/100)$
(Peterson~\cite{peterson}). The B-band magnitude was transformed
from the Galactic extinction corrected SDSS psf magnitude in the u
and g band following B=g+0.17(u-g)+0.11 (Jester \cite{Jester2005}).
The B-band absolute magnitude was calculated as $M_B=B-5\log
d_L-25-K(z)$, where $d_L$ is the luminosity distance and $K(z)$ is
the K-correction, $K(z)=2.5({\alpha}_{\nu}-1)\log(1+z)$
(Peterson~\cite{peterson}.)}
 $M_{\rm B}=-22.27$.
 Of the 84 objects qualifying quasar
luminosity, all have radio-loudness R $>10$ (defined as
R=$L_{5GHz}/L_{4400\AA}$, Kellermann et~al.~\cite{kel89}) and thus
are radio-loud. The remaining 29 are Seyfert\,I galaxies, i.e.\
broad line radio galaxies (BLRGs). As a demonstration,
Fig.\,\ref{fig:spec_blagn} shows examples of the spectra of a quasar
and a BLRG, respectively.

We note that among these BLAGNs two have the FWHM of $H\alpha$ or
$H\beta$ smaller than 2000\,\kmps, namely, J094857.31+002225.5,
J164442.53+261913.2. Their SDSS spectra reveal relatively strong Fe
II multiplet and weak [O III] emission ([O III]/$H\beta$ $<3$),
satisfying the conventional criteria of NLS1s (Osterbrock \&
Pogge~\cite{op85}). It should be noted that NLS1 AGNs with strong
radio emission are very rare, and were found to be present in small
numbers only recently (see e.g. Komossa et~al.~\cite{kom06}; Whalen
et~al.~\cite{wha06}; Yuan et al.~\cite{yuan08}). In fact,
J094857.31+002225.5 and J164442.53+261913.2 were included in the
radio-loud NLS1 sample of Yuan et al.~(\cite{yuan08}).
Very recently, the gamma-ray emission from relativistic jets
in J094857.31+002225.5 (PMN J0948+0022) has been detected by Fermi/LAT (Abdo
et~al.~\cite{abdo}).

\subsection{Narrow line radio galaxies}
For the remaining 14 emission-lined objects, no significant broad
components of the lines can be detected. Traditionally, they were
collectively referred to as NLRGs in previous works about their
identification. Among them, nine have [O III] emission line
equivalent widths smaller than 10\,\AA, and thus belong to the class
of so-called weak-line radio galaxies (WLRGs) as suggested by
Tadhunter et al.~(\cite{tad98}). In fact, galactic nuclei with
narrow emission line spectra are of heterogeneous types, including
Seyfert\,II nuclei, low-ionization nuclear emission regions (LINERs,
Heckman~\cite{h80}), as well as star-forming galaxies. The nature of
LINER is still a matter of debate, though recent studies tend to
support the idea that the energy source is predominately from
accretion on to a central black hole, as in normal AGNs (e.g. Ho et
al.~\cite{b11b}). In this paper, we consider LINERs to be a subclass
of AGNs, though they represent AGNs at a low  level of activity.

\begin{figure}
\begin{center}
 \includegraphics[width=0.33\hsize,height=0.33\hsize]{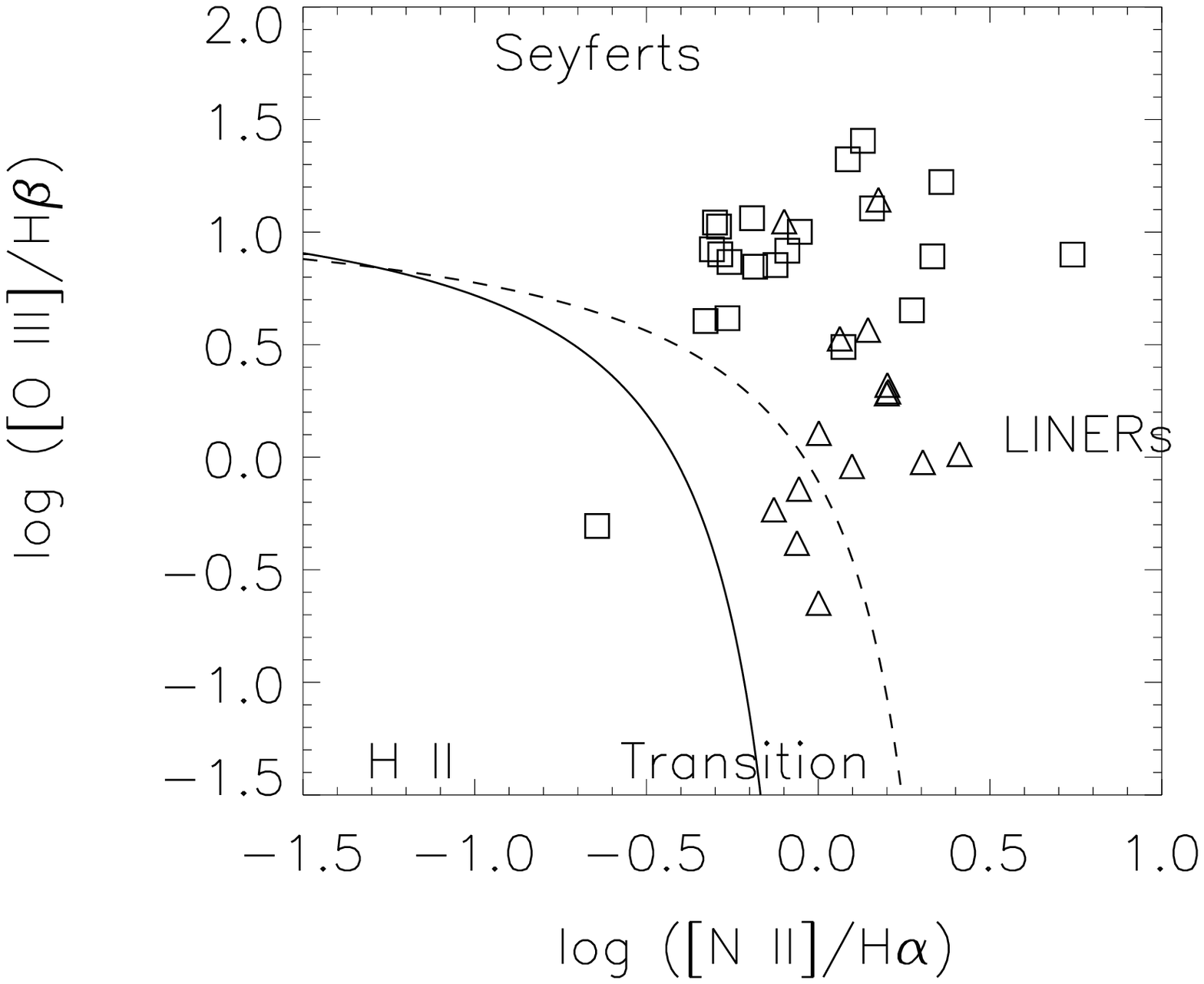}%
 \includegraphics[width=0.33\hsize,height=0.33\hsize]{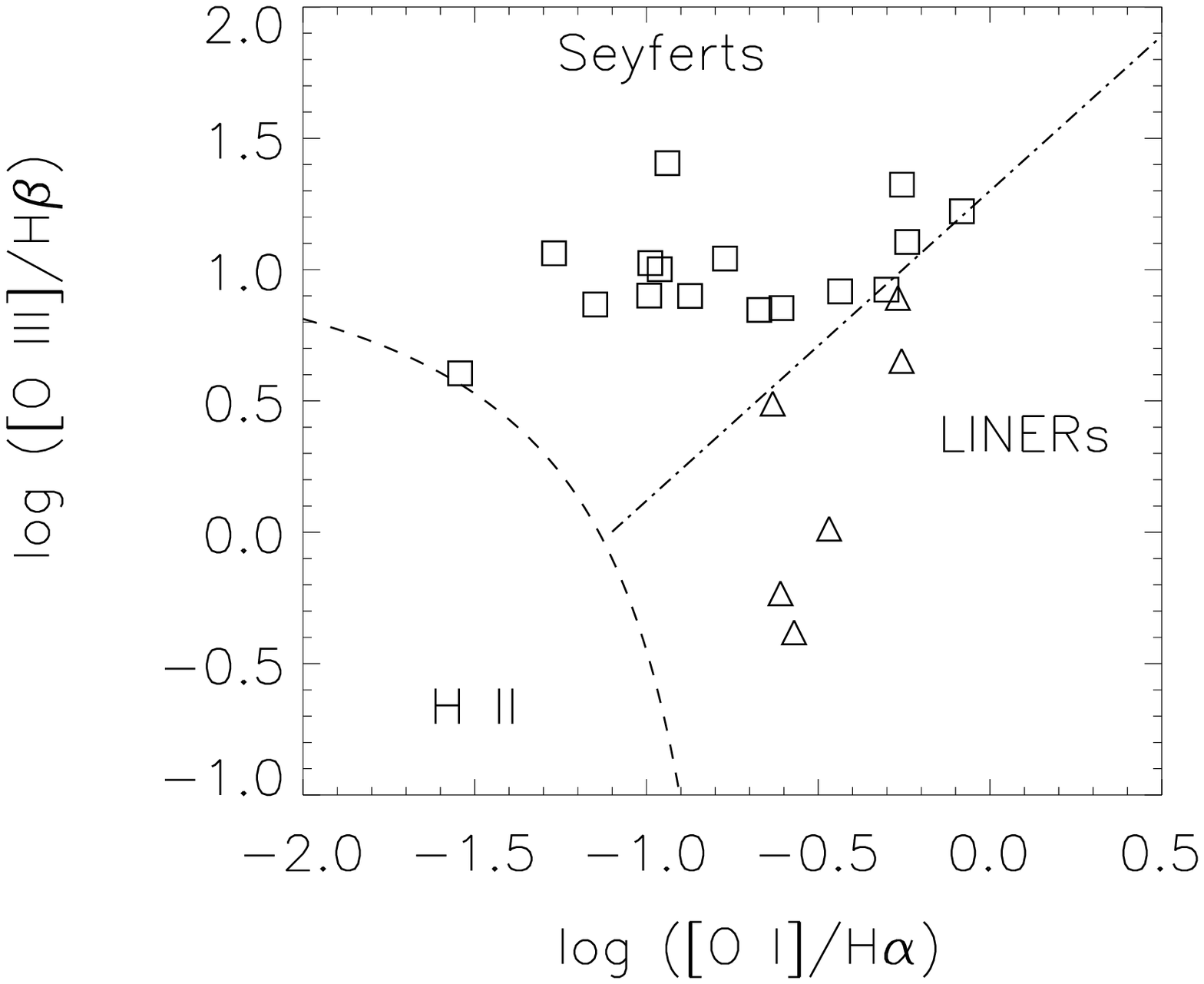}%
 \includegraphics[width=0.33\hsize,height=0.33\hsize]{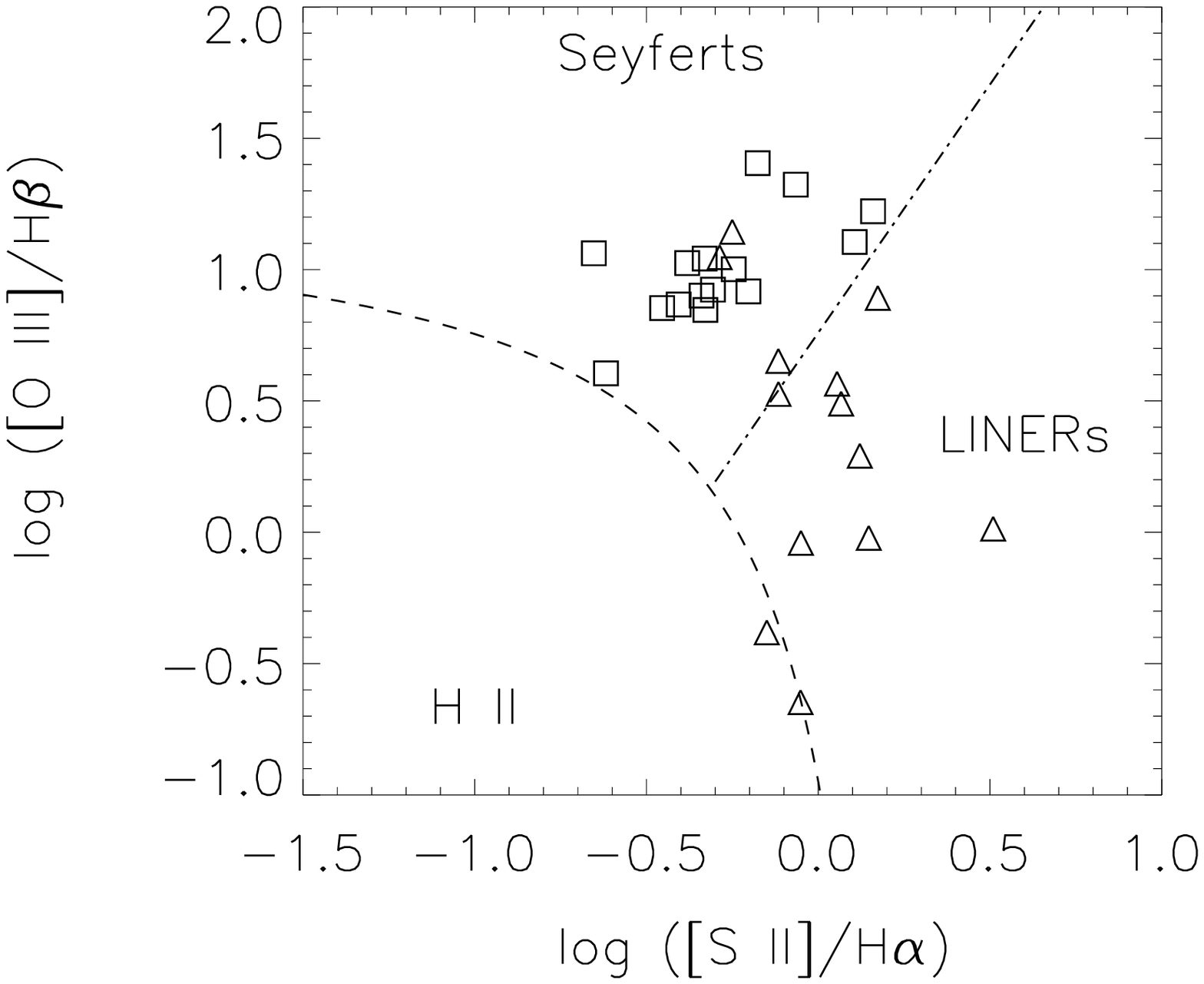}
\caption{\label{fig:lineratio} Diagnostic diagrams of emission line
ratios separating Seyfert galaxies, LINERs, and star-forming
galaxies for emission line objects in the RGB-SDSS sample. The
dividing schemes are adopted from Kauffmann et~al.~(\cite{k03}) and
Kewley et~al.~(\cite{k06}). The solid line separates star-forming
galaxies from Seyferts and LINERs (Kauffmann et~al.~\cite{k03}),
while the dashed line represents the maximum star-formation given by
Kewley et~al.~(\cite{k01}). The dashed-dotted line separates
Seyferts from LINERs (Kewley et~al.~\cite{k06}). Plot symbols:
squares for BLRG and triangles for NLRG. }
\end{center}
\end{figure}

\begin{figure}
\begin{center}
 \includegraphics[width=1.0\hsize,height=0.5\hsize]{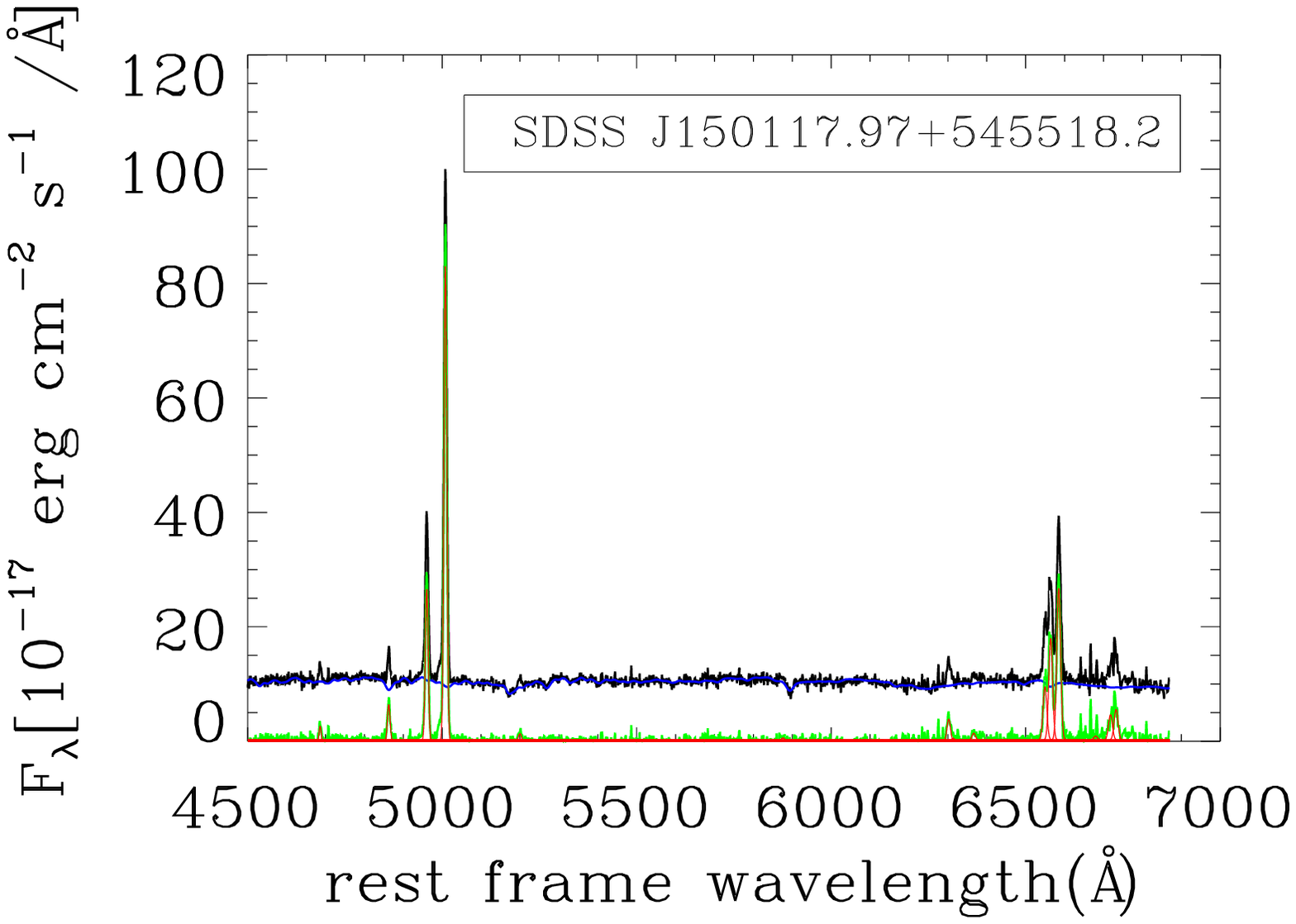}
 \includegraphics[width=1.0\hsize,height=0.5\hsize]{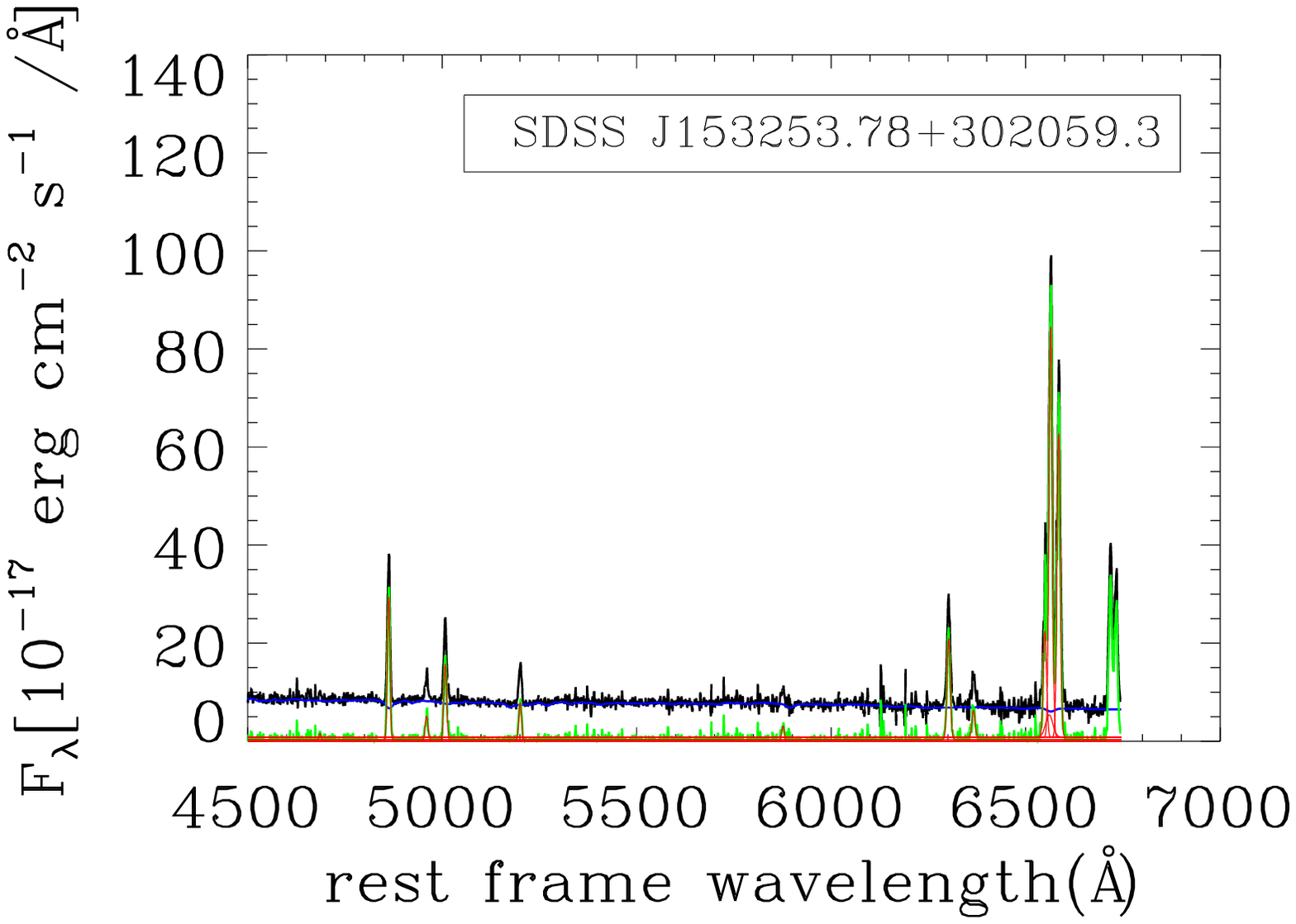}
 \caption{\label{fig:spec_nlrg}
Example spectra of identified narrow line AGNs in the rest frame for
a Seyfert\,II galaxy (upper panel) and a LINER (lower panel). See
Fig.\,\ref{fig:spec_blagn} for the plot coding.
 }
 \end{center}
\end{figure}

Further classification of the narrow emission line nuclei of
galaxies can be carried out based on a set of emission line ratios,
as suggested by Baldwin et~al.~(\cite{bpt}) and developed by a
number of authors (e.g. Veilleux \& Osterbrock~\cite{vo87}; Kewley
et~al.~\cite{k01}; Kauffmann et~al.~\cite{k03}; Kewley
et~al.~\cite{k06}). Here, we adopt the most recent variant of this
classification scheme as proposed by Kewley et~al.~(\cite{k06}) and
Kauffmann et~al.~(\cite{k03}). In this scheme, AGNs (Seyferts and
LINERs) are best separated from star-forming galaxies on the [N
II]/$H\alpha$ versus [O III]/$H\beta$ diagnostic diagram. Seyferts
and LINERs are further distinguished using the [O I]/$H\alpha$ and
[S II]/$H\alpha$ ratios, as shown in Fig.\,\ref{fig:lineratio}.
We found two Seyfert\,II and 12 typical LINER spectra(see
Fig.\,\ref{fig:spec_nlrg} for example spectra). No star-forming
galaxy is found, as expected for sources with strong X-ray and radio
emission as studied here. All of the nine WLRGs are classified as
LINERs. Moreover, we also examined the LINER classification using
the traditional criteria of [O II]$\lambda3727$/[O
III]$\lambda$5007$\geq1$ and [O I]$\lambda6300$/[O
III]$\lambda5007\geq1/3$ of Heckman~(\cite{h80}), and found that
these criteria were also met for most of the LINERs. It should be
noted that a few of the LINERs identified here show somewhat higher
ionization atypical of LINERs. In fact, these objects are located
very close to the border line separating LINERs and Seyferts in the
diagnostic diagram of Kewley et~al.~(\cite{k06}) and Kauffmann
et~al.~(\cite{k03}) and may have been classified as Seyferts if
other classification criteria were used, e.g. Veilleux \&
Osterbrock~(\cite{vo87}), Heckman~(\cite{h80}) and Ho
et~al.~(\cite{b11b}). Thus they are actually border-line objects
between Seyferts and LINERs. We further point out that the LINER
J151838.90+404500.2 was classified as Seyfert I galaxy by
Laurent-Muehleisen et~al.~(\cite{LM98}); however, no significant
broad component is found in its SDSS spectra.

\begin{figure}
\begin{center}
 \includegraphics[width=0.5\hsize,height=0.5\hsize]{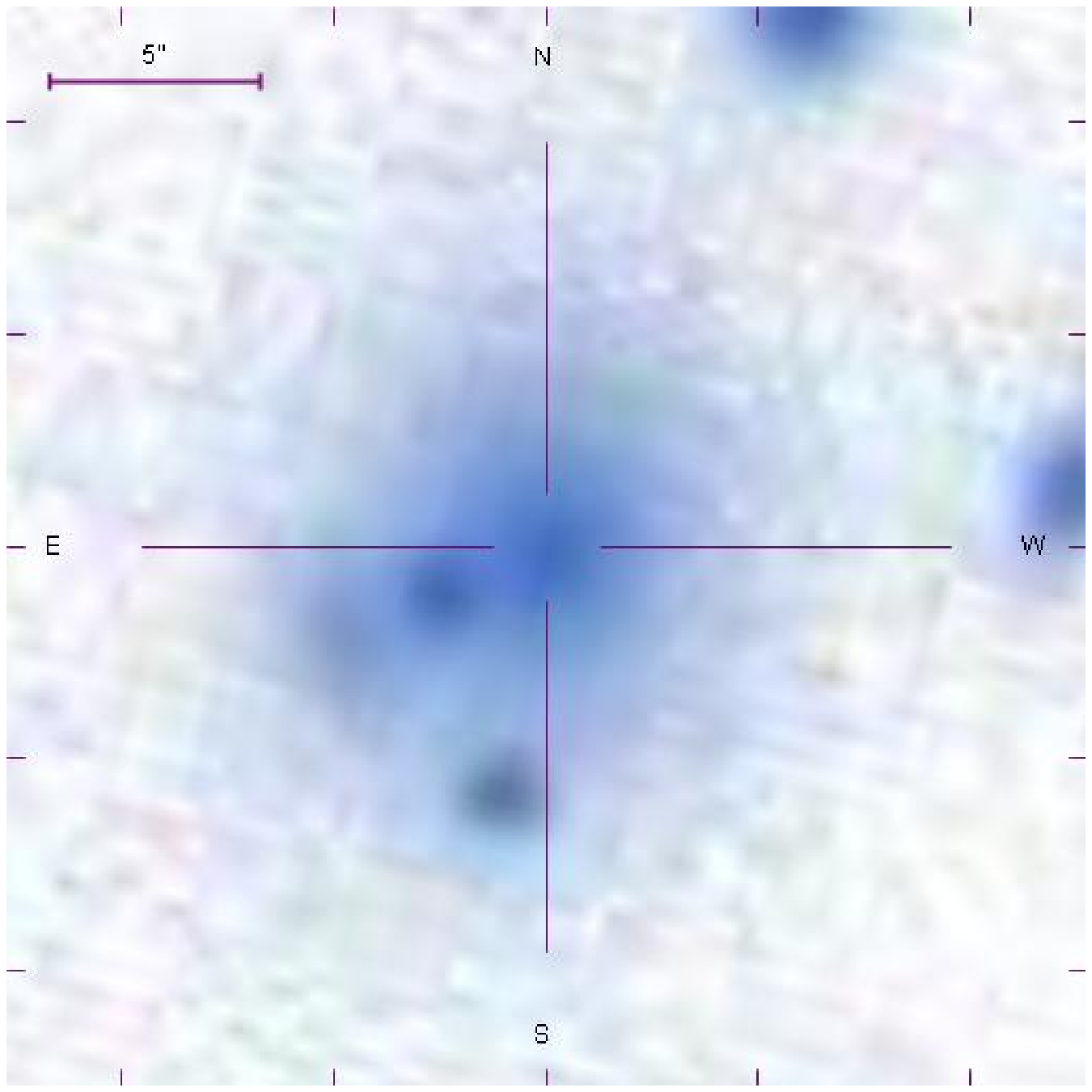}
 \caption{\label{fig:img_j1114}
Optical SDSS image of the LINER SDSS\,J111421.76+582319.8 (marked by
the cross) shows that it is in a merging system with one or more
other galaxies/cores. }
\end{center}
\end{figure}

We notice that the host galaxies of some sources identified with
LINERs are associated with clusters of galaxies, as given by NED
(Crawford et~al.~\cite{craw}; Koester et~al.~\cite{koes1}); these
are J111421.76+582319.8, J111908.94+090022.8, J113518.79+125311.1,
J132419.67+041907.0, J153253.78+302059.3, J160239.61+264606.0 and
J172010.03+263732.0. Their host galaxies are most likely giant
ellipticals in the centers of the clusters. For these objects the
X-ray fluxes are mostly the diffuse emission of the clusters. Three
of the objects, namely, J111421.76+582319.8, J153253.78+302059.3 and
J172010.03+263732.0,
 were previously observed in optical by  Crawford et~al.~(\cite{craw}),
 and the SDSS spectra and the narrow line ratios are consistent with
 those presented in Crawford et~al.~(\cite{craw}).
Of particular interest, J111421.76+582319.8 is in a merging system,
with two nuclei separated by 2.3$''$ at nearly the same redshift.
Its SDSS image is shown in Fig.\,\ref{fig:img_j1114}. Three of the
classified LINERs (J113518.79+125311.1, J150324.77+475829.6 and
J085004.65+403607.7) have the  CaII H \& K break values smaller than
0.40, indicating possible contribution of a non-thermal (BL Lac)
component in their optical emission, as previously found in
J150324.77+475829.6 (Edge et~al.~\cite{edge2003}).

\subsection{BL Lac objects and normal galaxies}
\begin{figure}
\begin{center}
 \includegraphics[width=1.0\hsize,height=0.5\hsize]{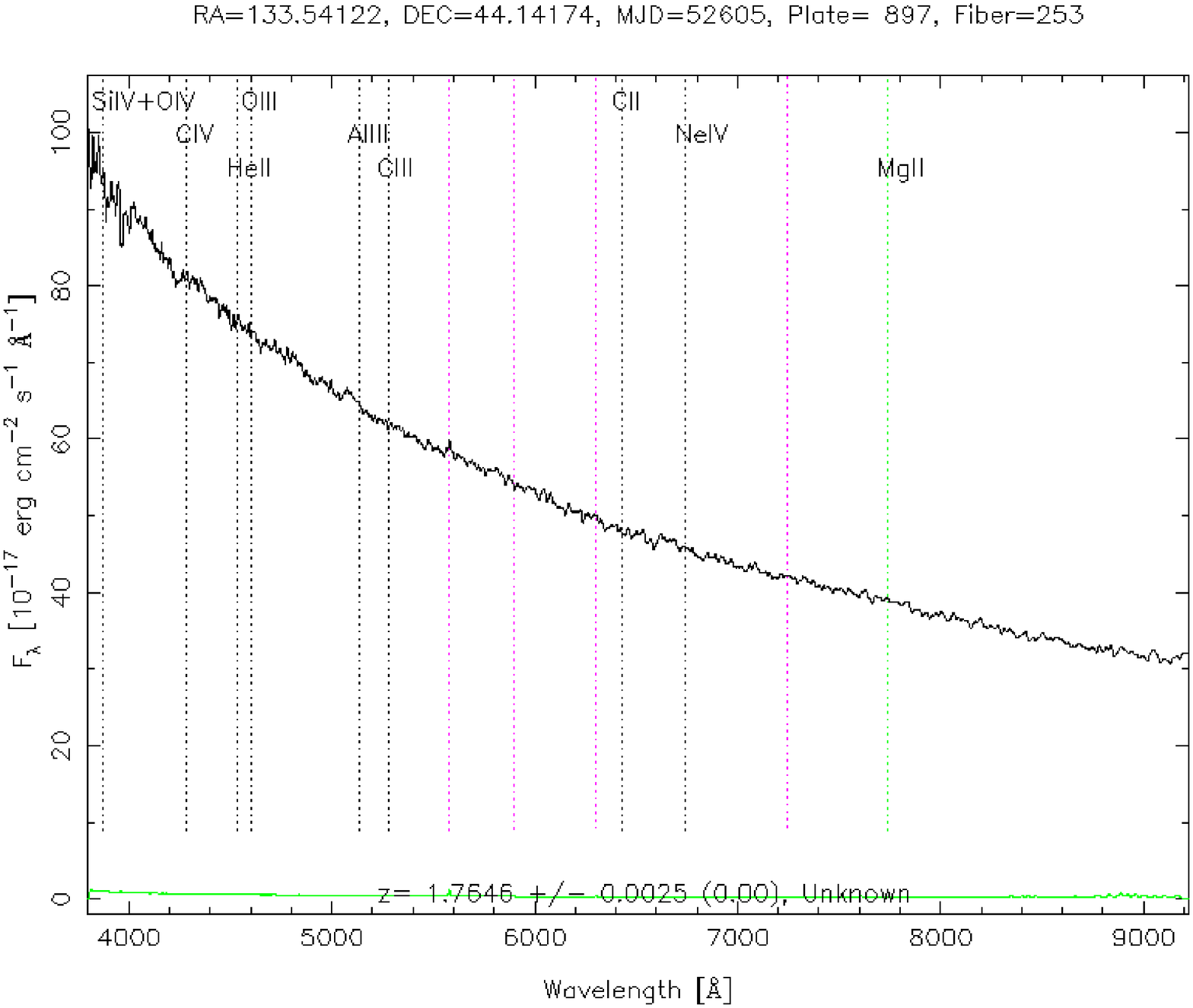}
 \caption{\label{fig:spec_bllac0}
  Example SDSS spectrum of a BL Lac object with a featureless continuum.
 }
 \end{center}
\end{figure}

\begin{figure}
\begin{center}
 \includegraphics[width=1.0\hsize,height=0.5\hsize]{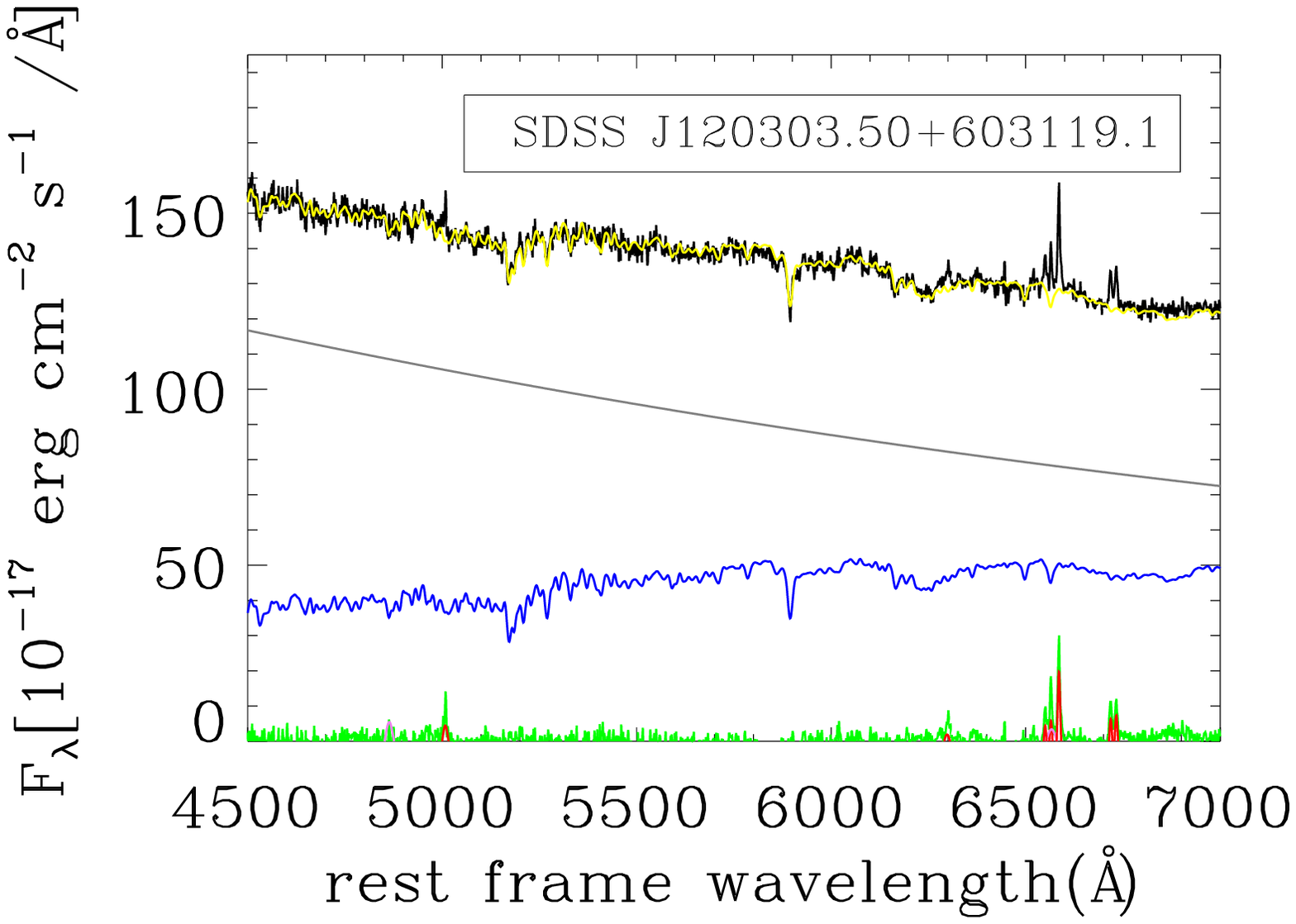}
 \includegraphics[width=1.0\hsize,height=0.5\hsize]{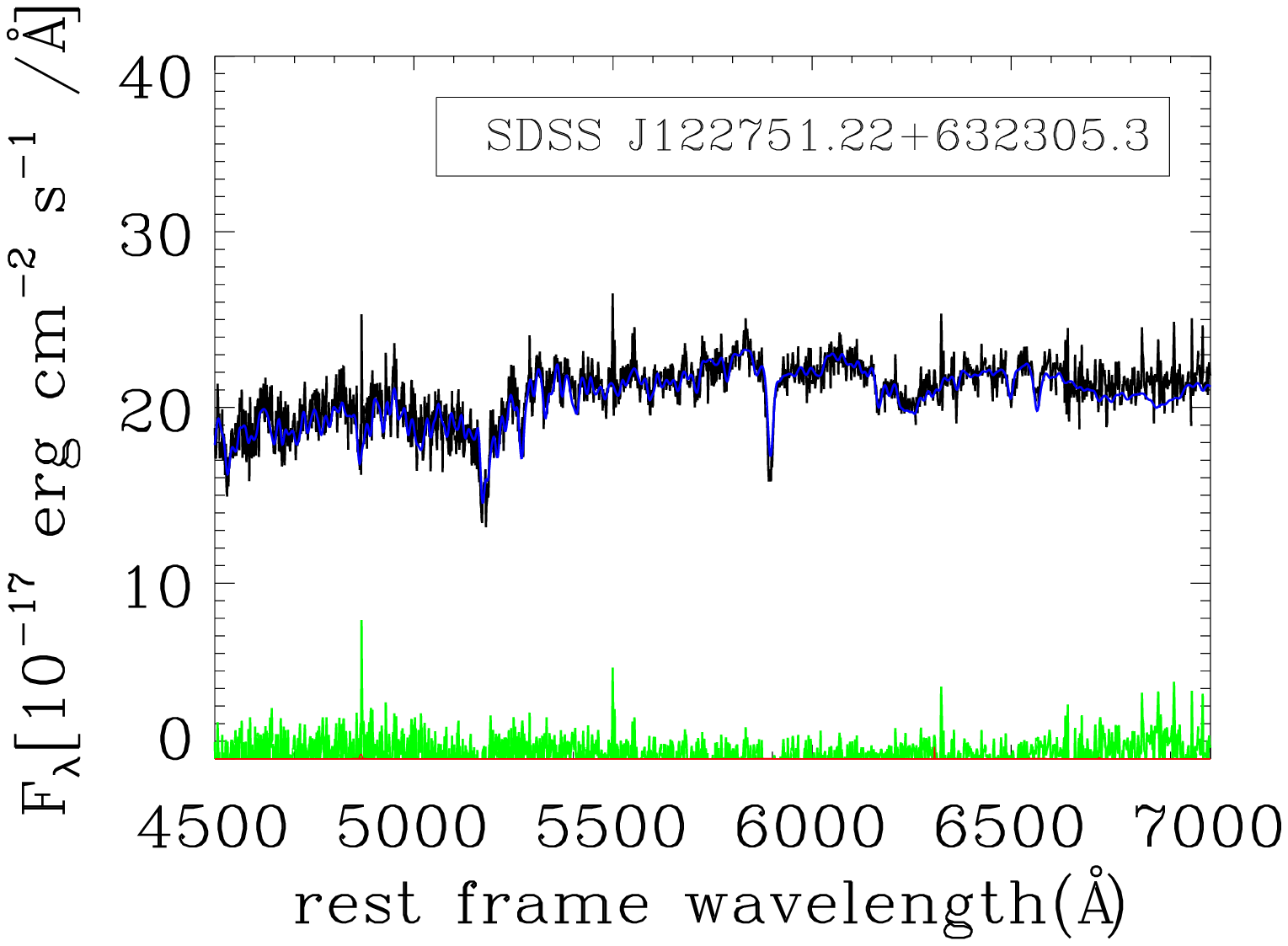}
 \caption{\label{fig:spec_bllac}
Example spectra of an identified weak-lined BL Lac object (upper
panel) and a normal galaxy (lower panel) in the rest frame. See
Fig.\,\ref{fig:spec_blagn} for plot coding.
 }
 \end{center}
\end{figure}

\begin{figure}
\begin{center}
 \includegraphics[width=0.5\hsize,height=0.5\hsize]{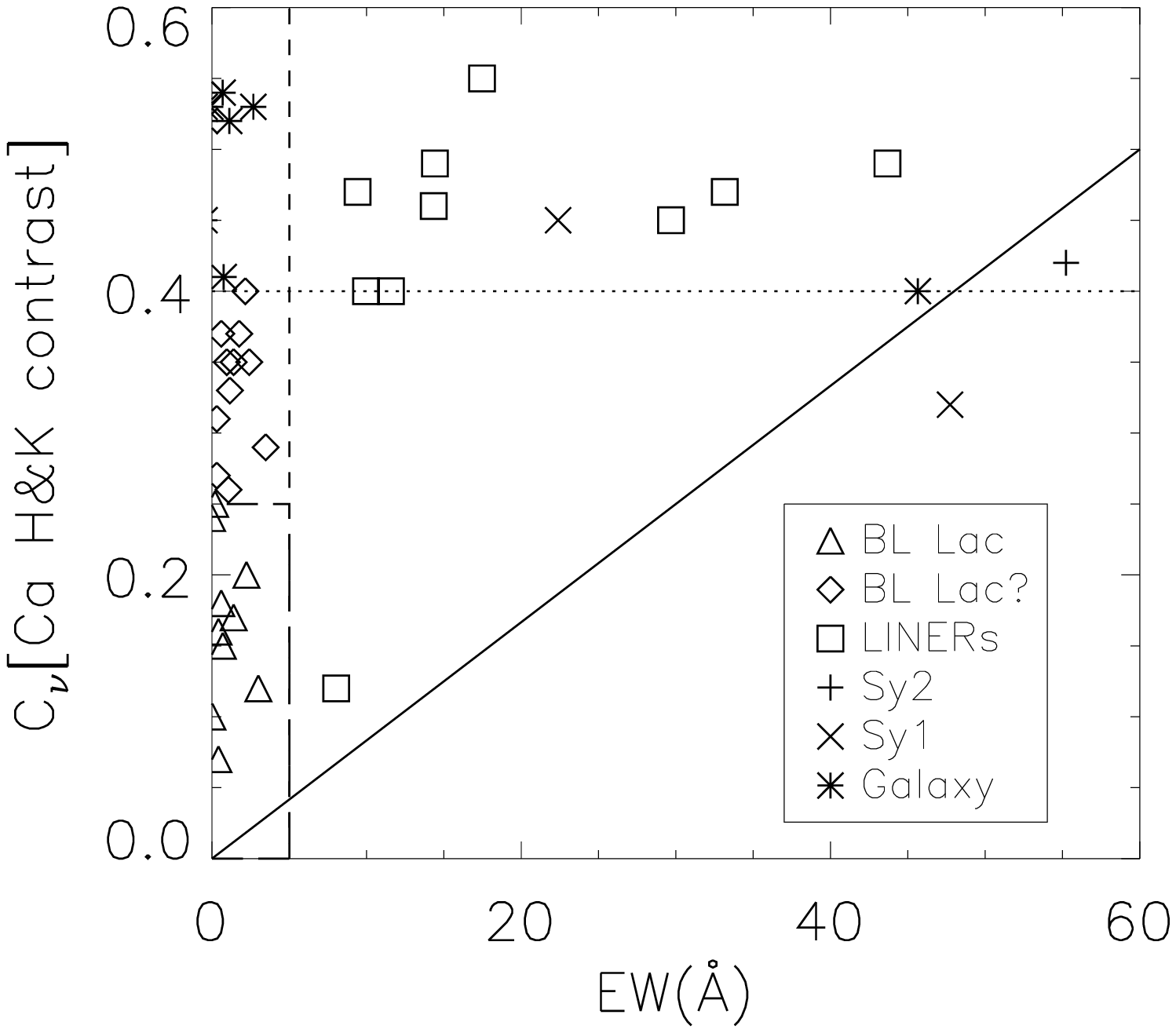}
 \caption{\label{fig_ew}
The Ca H\&K break contrast versus the EW of the strongest emission
line for objects without significant emission lines in the RGB-SDSS
sample. The EW distribution is only shown out to EW = 60\AA \,as in
Marcha et~al.~(\cite{Mar96}). The solid line represents the expected
decrease for the EW of an emission line in the presence of an
increasing amount of continuum following Marcha
et~al.~(\cite{Mar96}). The dotted line marks the Ca H\&K break
contrast $C_{\nu}$ = 0.40 and the short dashed line marks EW = 5\AA.
The area enclosed by the long dashed line corresponds to EW $\leq$
5\AA\,and Ca H\&K break contrast $\leq$ 0.25, as used by Stocke et
al.~(\cite{sto91}) for the classification of BL Lac. Plot symbols:
triangles for BL Lacs, diamonds for BL Lac candidates, squares for
LINERs, plus signs for Sy IIs, crosses for Sy Is, and asterisks for
galaxies. }
\end{center}
\end{figure}

The second group of objects, those without significant emission
lines, broadly show two kinds of spectra: a nearly featureless
continuum (see table \,\ref{bllac0} and see
Fig.\,\ref{fig:spec_bllac0} for example spectra) and a stellar
spectrum with absorption lines. In the case where a prominent
featureless continuum (unlike stellar spectrum) is present, we
attribute it to non-thermal emission from BL Lac objects (see
Fig.\,\ref{fig:spec_bllac} for example spectra). In case of the
absence of an apparent featureless continuum, we test its possible
presence against the starlight using the contrast of the CaII H \& K
absorption lines, i.e.\ the CaII H \& K break (Marcha
et~al.~\cite{Mar96}; Landt et~al.~\cite{L02}; Collinge
et~al.~\cite{Collinge05}). The CaII H \& K break is defined as
\[
C_{\nu}=0.14+0.86C_{\lambda}
\]
\[
C_{\lambda}=(f_{\lambda,+}-f_{\lambda,-})/f_{\lambda,+}
\]
where $f_{\lambda,-}$ and$f_{\lambda,+}$ are the average flux
densities in the rest-frame wavelength regions 3750-3950 and
4050-4250 \AA. This method is justified because the CaII H \& K
break of elliptical galaxies should always be grater than 0.4
(Dressler \& Shectman~\cite{ds87}). The inclusion of starlight in
the spectrum also dilutes the emission lines of AGNs. We adopt the
method suggested by Marcha et~al.~(\cite{Mar96}) to further
distinguish BL Lac objects and Seyfert galaxies by considering both
the CaII H \& K break and the equivalent width of emission lines, if
any exist. The criteria are illustrated in Fig.\,\ref{fig_ew}, as
from Marcha et~al.~(\cite{Mar96}), which shows the CaII H \& K break
versus the equivalent width of the strongest line. Using this
criteria, we classify objects with a CaII H \& K break greater than
40\% (without a significant non-thermal continuum) as `normal'
galaxies. Those with the CaII H \& K break $<$ 25\% and the rest
frame emission line $EW<$ 5\AA\ are classified as BL Lac objects;
and those with the break between 25\% and 40\% and $EW<$ 5\AA\ are
classified as BL Lac object candidates. Those with line $EW$s
marginally exceeding 5\AA\ are classified, depending the CaII H \& K
break, as either BL Lac objects candidates (smaller line EW) or
LINERs or Seyfert galaxies (larger line EW) (see fig\ref{fig_ew}).
As a result, we found 32 BL Lac objects, 11 BL Lac candidates, and
11 `optical normal' galaxies.

\section{Summary and discussion}

\subsection{Summary of optical identification}

The identification of the RGB-SDSS sample objects is summarized in
Table\,\ref{tab:spec_cls}. For 74 sources, the identifications are
presented for the first time. There are 84 radio-loud quasars (45
FSRQs), 27 BLRGs, 2 radio-loud NLS1 galaxies, 14 NLRGs comprising 12
LINERs and 2 Seyfert\,II galaxies), 32 BL Lac objects and 11 BL Lac
candidates, 11 `optical normal' galaxies (at least 6 in clusters).
Thus the vast majority of the RGB sources are identified with
radio-loud AGNs, particularly blazars, confirming previous results.
The advantage of our result is the refined spectroscopic
classification based on emission line ratios for NLRGs. Most of
these NLRGs have weak emission lines (WLRGs) and have typical
optical LINER spectra(12/14).

Of the RGB--SDSS sample, 36 sources have been previously identified
by Laurent-Muehleisen et~al.~(\cite{LM98}) using 2-m class
telescopes. Our results are in good agreement with theirs except for
two objects (see Table\,\ref{tab_comp}).
Furthermore, in the course of our work, Turriziani
et~al.~(\cite{Tur07}) published results of the identification of the
ROXA sample, which was compiled from the RASS and the NVSS radio
survey, and also made use of SDSS spectroscopic data. We note that
80 objects in that sample are included in our RGB--SDSS sample. For
most of them (69 source), our classifications are consistent with
those of Turriziani et~al.~(\cite{Tur07}), while this is not the
case for the remaining sources (11) , as listed in
Table\,\ref{tab_comp}. It should be pointed out that Turriziani
et~al.~(\cite{Tur07}) simply adopted the classification and spectral
parameters given by the SDSS pipeline, which are coarse in general
and even incorrect in some cases. We thus argue that our results are
more accurate and reliable, especially for AGNs with weak emission
lines or with narrower broad-lines than usual.

\subsection{Broad band SED}

From the radio, optical, and X-ray data of the sample objects, we
calculated their broadband effective spectral indices $\alpha_{ox}$,
$\alpha_{ro}$ and $\alpha_{rx}$ between 5\,GHz radio, 5000\,\AA\
optical, and 1\,keV X-ray using the monochromatic luminosities,
\[
 {\alpha}_{ox} = - \frac{\log (L_{1keV}/L_{5000\AA})}{\log ({\nu}_{1keV}/{\nu}_{5000\AA})}
 = - 0.384 \log (\frac{L_{1keV}}{L_{5000\AA}})
\]
\[
 {\alpha}_{ro} = - \frac{\log (L_{5000\AA}/L_{5GHz})}{\log ({\nu}_{5000\AA}/{\nu}_{5GHz})}
 = - 0.196 \log (\frac{L_{5000\AA}}{L_{5GHz}})
\]
\[
 {\alpha}_{rx} = - \frac{\log (L_{1keV}/L_{5GHz})}{\log ({\nu}_{1keV}/{\nu}_{5GHz})}
 = - 0.130 \log (\frac{L_{1keV}}{L_{5GHz}})
\]
 The choice of frequencies, as introduced in Ledden \&
Odell~(\cite{led85}), makes it easy to compare with the results of
Padovani et~al.~(\cite{P03}). The optical monochromatic fluxes at
5000\AA\ in the rest frame were derived from the Galactic extinction
corrected SDSS PSF magnitude in the i-band using the extinction map
of Schlegel et~al.~(\cite{Schlegel98}), assuming an optical spectral
index \alpoe=0.5 for the K-correction in the frequency domain. The
monochromatic X-ray fluxes were derived from the integrated
0.1--2.4\,keV fluxes corrected for Galactic absorption which are
given in Brinkmann et al.~(\cite{Brinkmann97}); for this, the X-ray
spectral indices were also taken from Brinkmann et
al.~(\cite{Brinkmann97}) if available (estimated from the hardness
ratios) or assumed to be \alpxe=1.2 otherwise. For the radio
spectral indices, the estimated 1.4--5\,GHz indices were used if
available; otherwise a slope of $0.5$ was assumed.

\begin{figure}
\begin{center}
 \includegraphics[width=0.7\hsize,height=0.7\hsize]{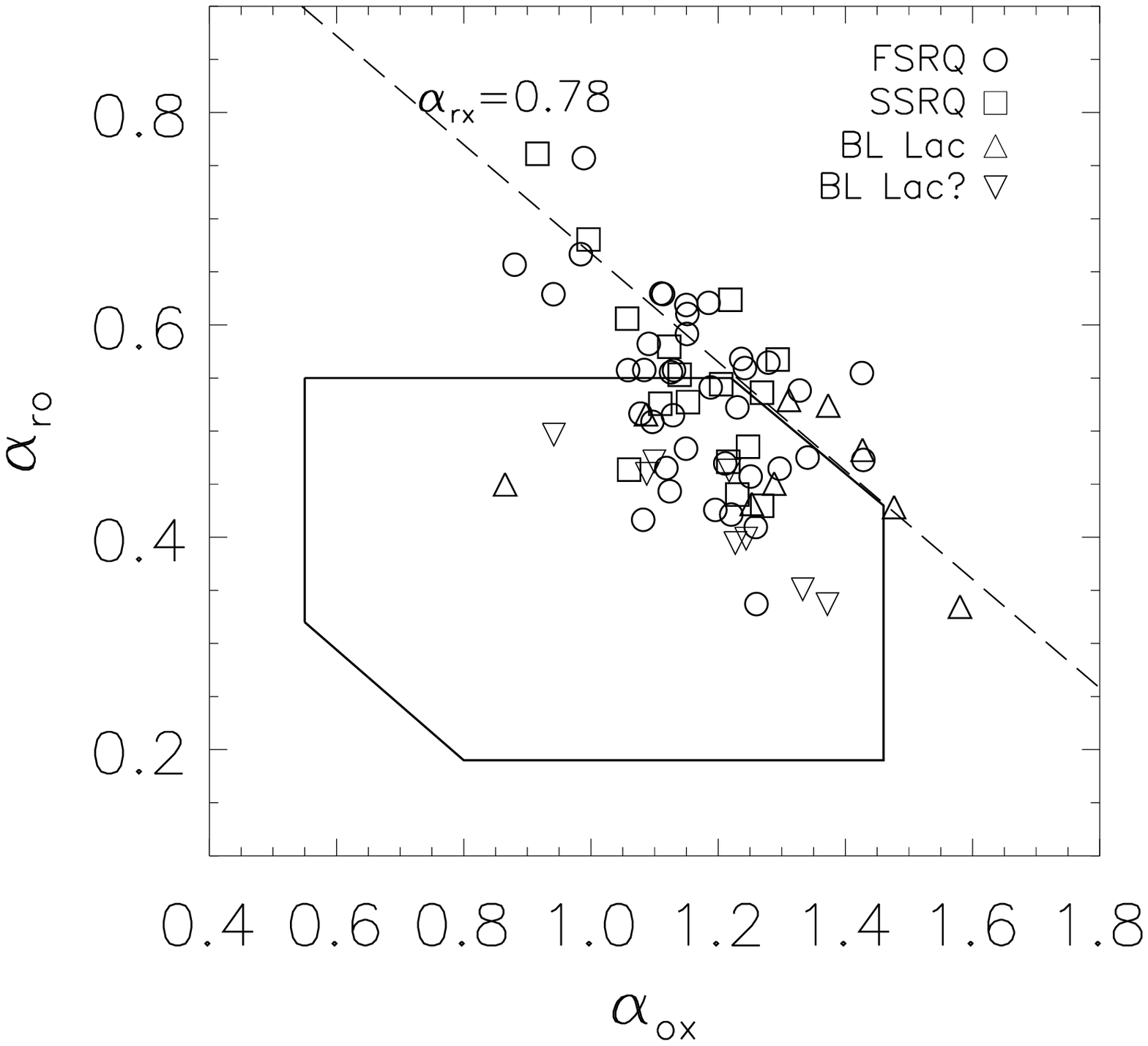}
 \caption{\label{fig_arox}
 \alpro versus \alpox for the identified radio-loud quasars and BL Lac
  objects in the RGB-SDSS sample. The dashed line represents a constant
   \alprx=0.78, which is commonlyused to divide BL Lac objects into
   LBLs and HBLs.A more conservative, schematic locus for HBL is also
   indicated following Padovani et~al.~(\cite{P03}) (solid line).
  Plot symbols: circles for FSRQs, squares for SSRQs, upside-down triangles
  for BL Lac candidates and triangles for BL Lac objects.
 }
 \end{center}
\end{figure}

Figure\,\ref{fig_arox} shows \alpro versus \alpox for the FSRQs and
SSRQs, as well as BL Lac objects identified in this work. It can be
seen that the distributions of the AGNs in the RGB--SDSS sample are
very consistent with those found in previous studies (e.g.\
Brinkmann et al.~\cite{Brinkmann95}, Padovani et al.~\cite{P03}) on
the \alpro--\alpox plane. The nominal dividing line between LBL
(low-energy peaked BL Lac objects) and HBL (dashed \alprxe=0.78) are
indicated (solid lines), as well as the more strict locus for HBL,
following Padovani et~al.~(\cite{P03}). The figure shows that while
many of the blazars (FSRQs and BL Lac objects) follow the trend of
LBL, there are indeed a number of objects falling within the locus
of HBL, including some FSRQs. Therefore, our result is consistent
with that of Padovani et~al.~(\cite{P03}) in the presence of  FSRQs
with the broad band two-point spectral indices resembling those of
HBL.

\begin{figure}
\begin{center}
 \includegraphics[width=0.5\hsize,height=0.5\hsize]{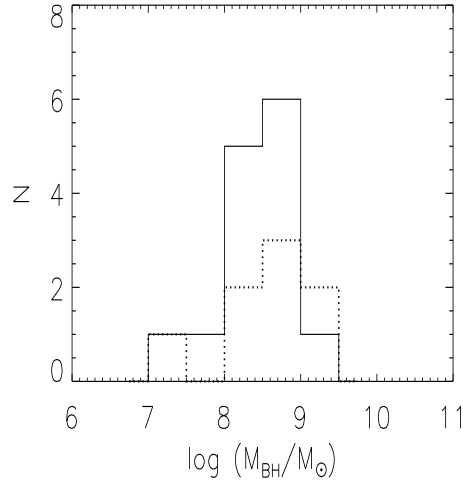}
 \caption{\label{fig_mbh}
   Distribution of the black hole masses for normal FSRQs (dashed line)
   and those with the two-point spectral indices similar to that of HBL (solid line).
 }
 \end{center}
\end{figure}

\begin{figure}
\begin{center}
  \includegraphics[width=0.5\hsize,height=0.5\hsize]{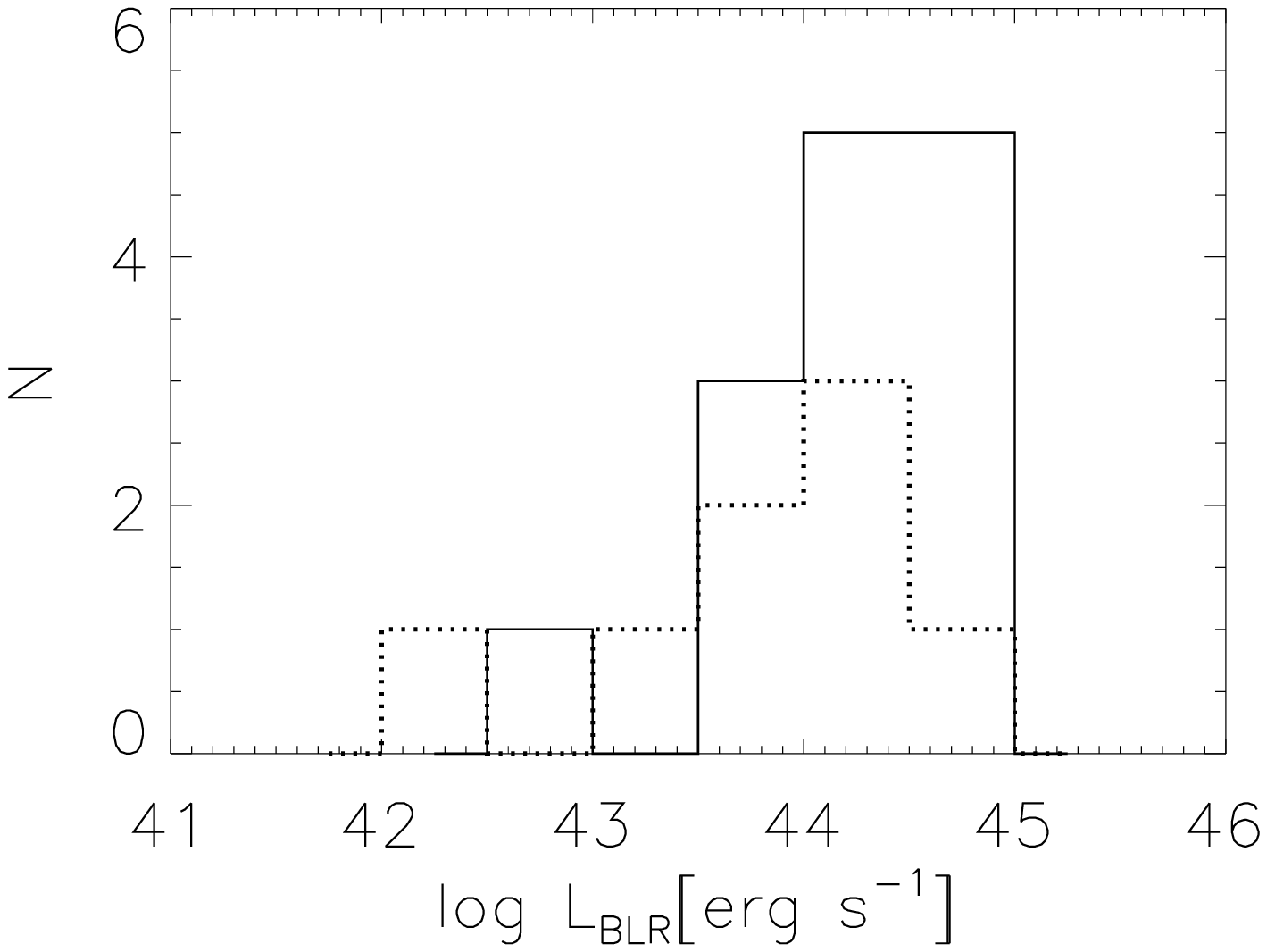}
  \caption{\label{fig_blr}
  Distribution  of the luminosities of the broad line region  for normal FSRQs (dashed line)
   and those with the two-point spectral indices similar to that of HBL (solid line)
  }
  \end{center}
\end{figure}

One would like to know if these unusual FSRQs differ from normal
FSRQs in terms of any other properties. We examined the
distributions of the central black hole mass (\mbh) and the
luminosity of the BLR (\lblr). The \mbh\ are estimated using the
width and luminosity of the broad \ha or \hb line, following
Vestergaard \& Peterson~(\cite{vp06}); Greene \& Ho~(\cite{gh07});
and McGill et~al.~(\cite{McG08}), for FSRQs with measured \ha or \hb
lines. The \lblr\ were calculated following the method suggested by
Celotti et~al.~(\cite{cel97}) \footnote{ Following Celotti
et~al.~(\cite{cel97}),
\[
L_{BLR}={\sum}_{i} L_{i,obs}\frac{<L_{BLR}^*>}{{\sum}_{i}
L_{i,ext}^{*}}
\]
where ${\sum}_{i} L_{i,obs}$ is the sum of the measured luminosities
of the observed broad lines, scaled by the ratio of the estimated
total luminosities $L_{BLR}^*$ to the estimated luminosities of the
observed broad lines. Both estimates were taken from the quasar
template spectrum of Francis et~al.~(\cite{fr91}) and
$H\alpha\,\lambda 6563$ from Gaskell et~al.~(\cite{gas81}). From
this approach, $L_{BLR}=25.26L_{H\beta}$ or $L_{BLR}=16.35L_{Mg II}$
or $L_{BLR}=7.22L_{H\alpha}$}. The distributions of \mbh\ and \lblr\
for both the HBL-like FSRQs and the classical FSRQs are shown in
Fig.\,\ref{fig_mbh} and Fig.\,\ref{fig_blr}, respectively. As can be
seen, and also confirmed by the K-S (Kolmogorov-Smirnov) test, the
\mbh\ and \lblr\ distributions are indistinguishable for these two
groups of FSRQs. We also compared the distributions of the Eddington
ratios \footnote{The bolometric luminosities $L_{bol}$ were
calculated following the relationship of Richards et
al.~(\cite{r06}) from the continuum luminosity.
 \[
 L_{bol}=9.26L_{5100\AA}
 \]
In order to eliminate the possible contamination of a non-thermal
emission component from relativistic jets in FSRQs, we estimated the
continuum luminosity from the emission line luminosity following the
relations in Greene \& Ho~(\cite{gh05}).} $L_{bol}/L_{edd}$ for the
HBL-like FSRQs (median 0.105) and classical FSRQs (median 0.029) and
concluded that they are marginally indistinguishable (with a
probability level of 0.03 by the K-S test) .

 %%% here .......

\begin{figure}
\begin{center}
  \includegraphics[width=0.6\hsize,height=0.5\hsize]{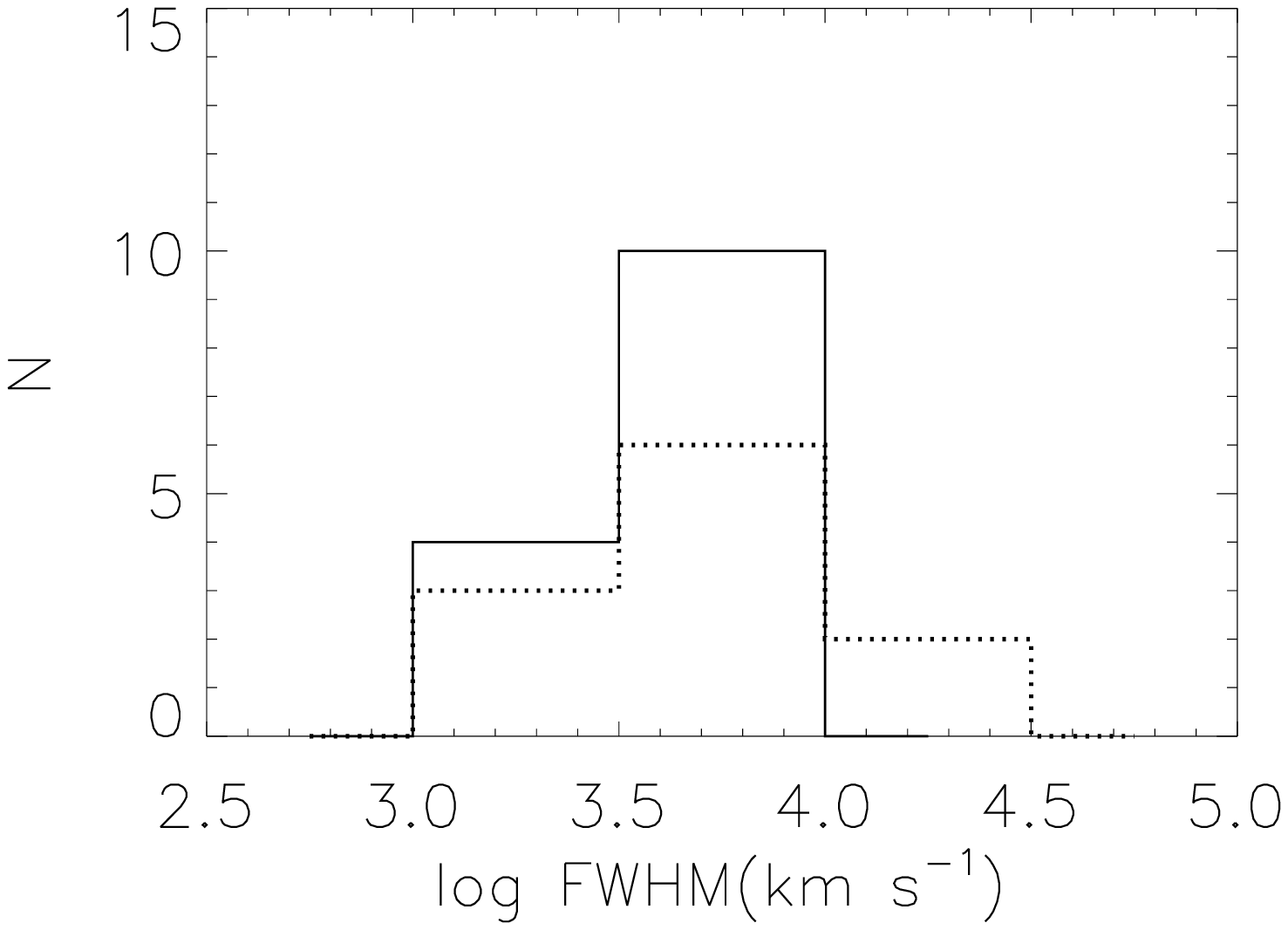}
  \caption{\label{fig_fwhm}
   Distribution of the broad line width (FWHM) for normal FSRQs (dashed line)
   and those with the two-point spectral indices similar to that of HBL (solid ).
  }
\end{center}
\end{figure}

In light of the finding of Yuan et al.\ (2008) that a fraction of
the very radio-loud NLS1 AGNs in their sample has an HBL-like
\alproe--\alpox distribution, it would be interesting to ask whether
there is a link between NLS1s (or the narrowness of broad emission
lines) and FSRQ with HBL-like SED. Figure\,\ref{fig_fwhm} shows the
line-width distribution of the broad lines (\hae, \hbe) for the
normal FSRQs including Seyferts (dashed line) and the 14 candidate
HBL-like FSRQs (solid line). The median line width of the HBL-like
FSRQs is 3699\,\kmpse, smaller than 6159\,\kmps for normal FSRQs and
Seyferts. However, the sample size is too small to confirm the
difference; the K-S test gives a probability level of 0.34. Thus,
our sample is too small to give a conclusive answer to this
question.

However, it should be pointed out that whether or not the broad band
spectral indices are good indicators of HBL-type blazars is still a
matter of active debate (for a different view, see Maraschi et
al.~(\cite{mar08})). Therefore, the determination of whether these
objects are genuine HBL-type FSRQs has to await proper estimation of
the synchrotron peak frequency by obtaining the broad band SED data
in the future.

\subsection{LINERs in the RGB-SDSS sample}

In previous identification work performed by Laurent-Muehleisen
et~al.~(\cite{LM98}) and also by some others, objects without broad
permitted emission lines were generally classified as NLRGs. Here,
we are able to break this degeneracy and further classify them into
Seyfert\,II nuclei and LINERs, based on their emission line ratios.
Of particular interest, our results show that the majority of the
narrow emission-line galaxies are identified with LINERs, and only a
small fraction are identified with Seyfert\,II nuclei. Furthermore,
all of those with weak emission lines (WLRGs) are in fact LINERs.
This result is consistent with that of Lewis et al.~(\cite{lew03}),
who found that a large fraction ($\sim50\%-75\%$) of WLRGs is in
fact LINERs, whenever a high quality optical spectrum was obtained.

\begin{figure}
\begin{center}
  \includegraphics[width=0.5\hsize,height=0.5\hsize]{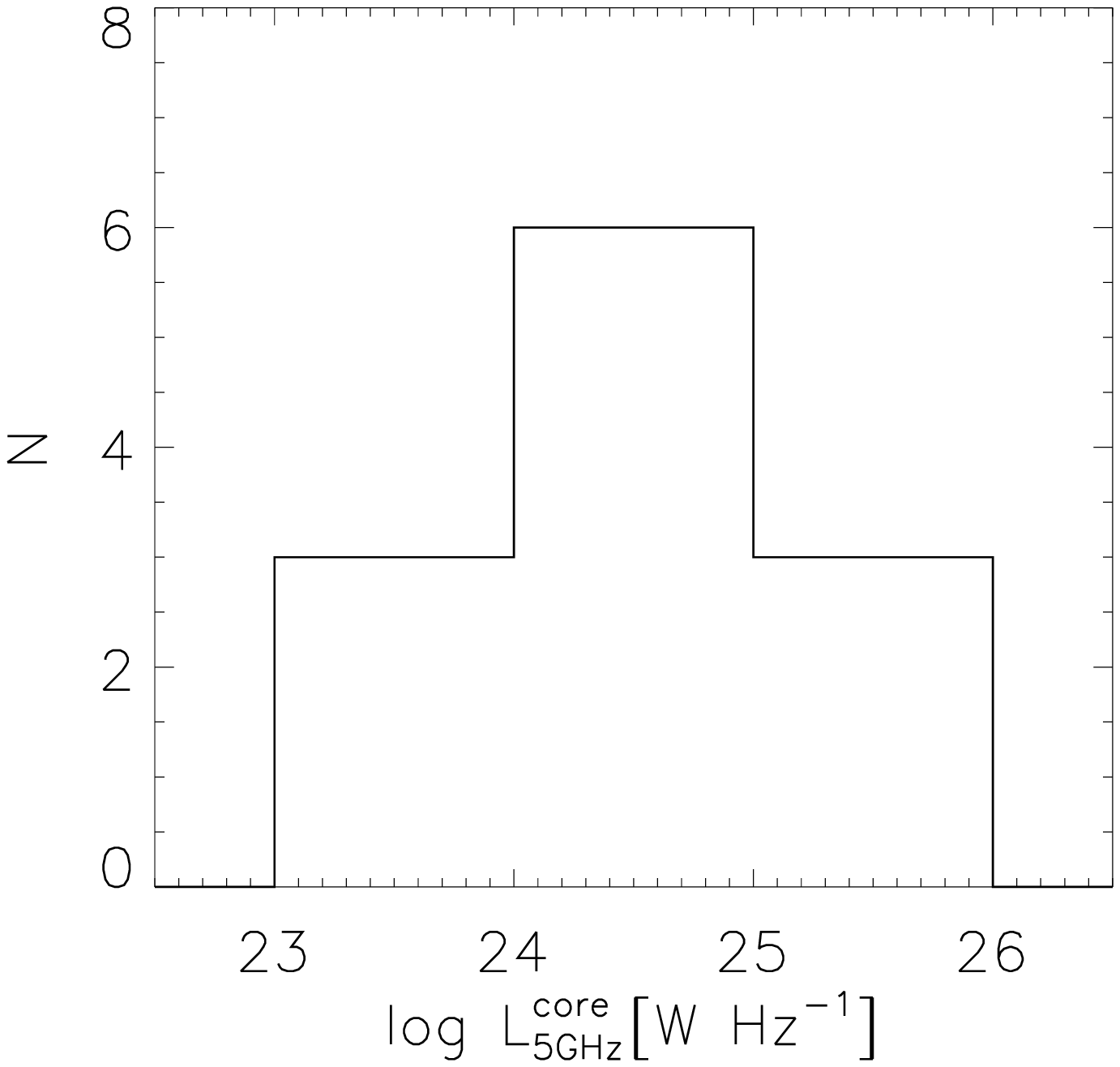}
  \caption{\label{fig_lglr}
  Distribution of the radio core luminosities for the LINERs identified
  in this work.}
\end{center}
\end{figure}

The nature of LINERs is not yet well understood, but recent studies
tend to suggest that a significant fraction of them consists of
low-luminosity AGNs (LLAGNs), i.e.\ powered by black hole accretion
(e.g. see Ho~(\cite{ho08}) for a recent review). The radio
properties of LINERs have been studied extensively over the years.
It has been established that they often show radio core emission
(e.g. Heckman~\cite{h80}; Nagar et al.~\cite{nag05}) or sometimes
even sub-parsec-scale jets (e.g. Falcke et al.~\cite{fal00}; Nagar
et al.~\cite{nag05}), mostly with flat or event inverted spectra
(e.g. Nagar et al.~\cite{nag00}; Falcke et al.~\cite{fal00}). A
detection rate of compact radio cores as high as $44\%$ was found in
LINERs, similar to that in Seyferts (Nagar et al.~\cite{nag05}).
Given that LINERS are radio emitters with flat spectra, the
identification of some of the RGB sources with LINER is not
surprising. However, for the majority of LINERs,  the radio powers
are generally quite modest---mostly in the range of
$10^{19-21}$\,\whz\ at 5\,GHz for optically selected samples. The
modest radio power is consistent with their low nuclear
luminosities, in light of the suggested correlation between radio
luminosity and emission line luminosity (e.g. Nagar et
al.~\cite{nag05}).
In comparison, some of the LINERs identified in this work have very
high radio core luminosities, \lrc=$10^{23}-10^{25}$\,\whz at 5\,GHz
(see Fig\,\ref{fig_lglr}). They are of particular interest as they
are possibly among the strongest radio emitters of the LINER
population. In fact, these objects are at the higher-end of the
radio luminosity function either of E/S0 galaxies (Sadler et
al.~\cite{sad89}) or of LLAGNs (Nagar et al.~\cite{nag05}), with
both classes composed significantly of LINERs. We do not expect a
significant relativistic beaming effect in these objects, since
their radio core-dominance parameters are low or modest, as given in
Laurent-Muehleisen et~al.~(\cite{LM98}). Thus, the large radio
luminosities should  be intrinsic.

\begin{figure}
\begin{center}
 \includegraphics[width=0.5\hsize,height=0.5\hsize]{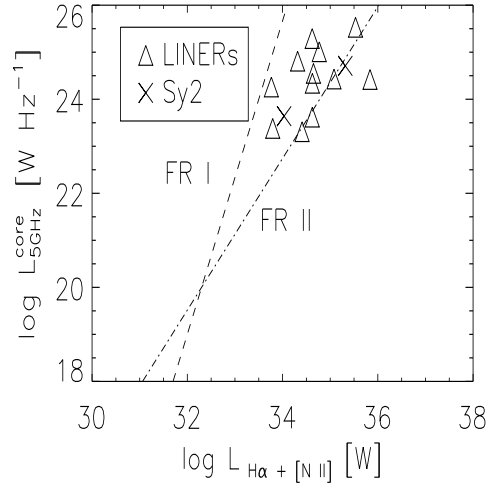}
 \caption{\label{fig:lrlha}
Radio core luminosity versus narrow emission line $H\alpha$+$[N II]$
$\lambda \lambda 6548,6583$ luminosity for the LINERs (triangles)
and Seyfert\,II galaxies (crosses) identified in this work. Also,
plotted are the linear fits of the same relationship for
FR\,I(dashed line) and FR\,II (dot-dashed line) radio galaxies from
Zirbel \& Baum ~(\cite{zb95}); see also Nagar et al.~(\cite{nag05})
}
\end{center}
\end{figure}

It has been suggested that the radio power \lr\ of LINERs is
correlated with the emission line luminosity (e.g. Nagar et
al.~\cite{nag05}). Indeed, the emission line luminosities (or the
power of the NLR) of these LINERs are very high, $L_{\rm H\alpha + N
II}$ ranging from $10^{40.7}$ to $10^{42.8}$\,\ulume, on average
$\sim2$ orders of magnitude higher than those of the LINERs in the
Palomar nearby galaxy sample of Ho et al.~(\cite{ho95}) which were
detected in radio by Nagar et al.~(\cite{nag05}). We estimated their
NLR luminosities following Celotti \& Fabian~(\cite{cel93})
\footnote{$L_{NLR}=3 (3 L_{[OII]3727} + 1.5 L_{[OIII]5007})$}, which
have $\log L_{\rm NLR}$(\ulume)=41.7--43.6, with a median of 42.2.
Interestingly, these values fall in between (with considerable
overlaps) those of high-power FRI/FRII radio galaxies and normal
low-power LLAGNs as found in the Ho et al.~(\cite{ho95}) sample and
other optically selected LINER samples. Hence, they are the radio
luminosities. This result is in line with the previous finding that
LINERs and low-luminosity Seyferts follow the same
radio--NLR-luminosity relation as FR\,I and FR\,II radio galaxies
(e.g. Nagar et al.~\cite{nag05}), which is shown in
Fig.\,\ref{fig:lrlha}. It can be seen that our LINERs are located
close to FR\,IIs, similar to the LINERs in non-elliptical host
galaxies in the Palomar sample (see the left panel of their Fig.\,3
in  Nagar et al.~(\cite{nag05})).

A more fundamental relationship is the dependence of the radio power
on both \mbh\,and the Eddington ratio, as suggested in e.g. Nagar et
al.~(\cite{nag05}). If this is the case, we anticipate that our
LINERs must have the highest possible accretion rates for LINER and
are harboring very massive central black holes. Among the 12 LINERs,
8 have reliable estimates of stellar velocity dispersion
($\sigma_*$) from the central galactic stellar spectra within the
3$''$ SDSS fibers, obtained via the nuclear-starlight decomposition
algorithm (Appendix\,A). We estimated the central black hole mass
using the well known \mbh--$\sigma$ relation (Tremaine
et~al.~\cite{trem02}), which ranges from $\log$\,(\mbh/\msune)= 8.4
-- 9.7 with a median of 9.3, as listed in Table\,\ref{liner}.
Indeed, our LINERs are among those galaxies harboring the most
massive central black holes, since this $M_{BH}$ range is in the
high-end of the Black Hole mass function of galaxies at low
redshifts(z $<$ 0.3) (McLure \& Dunlop~\cite{Mcl2004}, Greene \&
Ho~\cite{greene07}). Detailed investigation on the Eddington ratios,
radio jet power, as well as other multi-wavelength properties is
beyond the scope of this paper and will be pursued in future work.

The finding of radio luminous LINERs in this work is not surprising,
given the relatively high cutoffs in the radio/X-ray selection of
the RGB sample. We also compared our results with previous work in
the literature. We calculated the radio luminosities of the WLRGs
identified from the  2\,Jy sample (Tadhunter et al.~\cite{tad98})
which were further classified as LINERs by Lewis et
al.~(\cite{lew03}), and found that they have the same luminosity
range as our LINERs. We therefore suggest that an effective way of
finding powerful LINERs is to search for the associated LINERs with
strong radio sources.

%\section{Summary}

\section*{Acknowledgments}
This work is supported by the Chinese Natural Science Foundation
through project No.\ NSF10533050, and by the National Basic Research
Program of China - 973 Program 2009CB824800. Funding for the
creation and the distribution of the SDSS Archive has been provided
by the Alfred P. Sloan Foundation, the Participating Institutions,
the National Aeronautics and Space Administration, the National
Science Foundation, the U.S. Department of Energy, the Japanese
Monbukagakusho, and the Max Planck Society. The SDSS is managed by
the Astrophysical Research Consortium (ARC) for the Participating
Institutions. The Participating Institutions are the University of
Chicago, Fermilab, the Institute for Advanced Study, the Japan
Participation Group, the Johns Hopkins University, Los Alamos
National Laboratory, the Max-Planck-Institute for Astronomy (MPIA),
the Max-Planck-Institute for Astrophysics (MPA), New Mexico State
University, Princeton University, the United States Naval
Observatory, and the University of Washington. This research has
made use of the NASA/IPAC Extragalactic Database (NED) which is
operated by the Jet Propulsion Laboratory, California Institute of
Technology, under contract with the National Aeronautics and Space
Administration.

\begin{landscape}
\begin{center}
\begin{longtable}{ccccccccccccl}
\caption{RGB-SDSS Sample Objects and Identification}\\
\label{tab:spec_cls}\\
\hline
SDSS name&z&${\Delta}_{ro}$&$F_{6cm}$&$F_{20cm}$&$\alpha_{r}$&$u$&$g$&$i$&$M_{B}$&$\lg R$&SpecClass&ref\\
(1)&(2)&(3)&(4)&(5)&(6)&(7)&(8)&(9)&(10)&(11)&(12)&(13)\\
\hline
\endfirsthead
\hline
SDSS name&z&${\Delta}_{ro}$&$F_{6cm}$&$F_{20cm}$&$\alpha_{r}$&$u$&$g$&$i$&$M_{B}$&$\lg R$&SpecClass&ref\\
(1)&(2)&(3)&(4)&(5)&(6)&(7)&(8)&(9)&(10)&(11)&(12)&(13)\\
\hline
\endhead
\hline
J000132.83+145608.0& 0.3988& 0.04& 159/156& 314.6& 0.56&18.99&19.03&18.95& -22.00& 3.34&    Sy1&2\\
J011354.50+132452.4& 0.6853& 1.10& 148/150& 247.8& 0.40&18.67&18.38&18.52& -23.70& 3.09&   QSO&2\\
J073320.84+390505.2& 0.6637& 0.09&  86/100& 150.5& 0.33&18.18&18.02&18.10& -24.02& 2.70&   QSO&\\
J074906.50+451033.9& 0.1921& 0.17&  50/---& -----&  ---&16.55&16.72&16.25& -22.77& 1.88&    QSO&1\\
J075244.19+455657.3& 0.0518& 0.19&  59/---& -----&  ---&18.49&17.07&15.92& -19.31& 2.06&    Sy1&\\
J075448.86+303355.1& 0.7956& 0.12& 108/---& -----&  ---&17.72&17.32&17.42& -25.04& 2.54&    QSO&2\\
J080131.96+473616.0& 0.1569& 2.84&  41/---& -----&  ---&15.89&16.05&15.59& -23.00& 1.52&    QSO&1,2\\
J080644.42+484149.2& 0.3695& 0.17&  81/---& -----&  ---&17.59&17.54&17.55& -23.31& 2.45&    QSO&\\
J081058.99+413402.7& 0.5066& 0.10& 186/182& 219.7& 0.15&18.56&18.42&18.37& -23.07& 3.18&   QSO&2\\
J081100.60+571412.5& 0.6110& 0.15& 305/375& 460.0& 0.16&18.08&17.65&17.68& -24.18& 3.10&   QSO&\\
J081432.11+560956.6& 0.5093& 0.11&  47/ 43&  60.3& 0.27&18.05&17.90&17.90& -23.60& 2.37&   QSO&1\\
J081902.32+322637.2& 0.6512& 1.06& 274/209& 194.4&-0.06&21.36&20.75&19.72& -21.17& 4.30&    Sy1&\\
J081916.61+264203.2& 0.5261& 0.03&   7/117& 254.6& 0.63&18.93&18.61&18.66& -22.93& 1.83&   QSO&\\
J082814.20+415351.9& 0.2262& 0.21&   6/ 47&  90.5& 0.53&19.62&19.01&17.89& -20.70& 1.88&BL Lac?&2\\
J083353.88+422401.8& 0.2491& 0.33& 306/390& 248.6&-0.36&17.70&17.21&16.46& -22.73& 2.87& BL Lac&2\\
J084203.73+401831.3& 0.1516& 0.08&  15/---& -----&  ---&16.89&17.07&16.49& -21.91& 1.49&    Sy1&2\\
J084650.00+064149.0& 0.6161& 0.09&  62/103&  86.7&-0.14&17.48&17.08&17.20& -24.77& 2.18&   QSO&2\\
J085004.65+403607.7& 0.2667& 0.09& 112/110& 139.3& 0.19&20.98&19.75&18.22& -20.21& 3.45& LINERs&\\
J085036.20+345522.6& 0.1450& 0.27&  28/ 20&  34.9& 0.45&18.90&18.09&17.16& -20.62& 2.17&BL Lac?&\\
J085039.95+543753.3& 0.3673& 0.09&   6/---& -----&  ---&18.77&18.53&18.23& -22.27& 1.71&    Sy1&\\
J085348.18+065447.1& 0.2232& 0.12& 220/---& -----&  ---&20.30&19.75&18.41& -19.94& 3.74&    Sy1&\\
J085830.61+080422.8& 0.4549& 1.93&  31/---& -----&  ---&18.77&18.45&18.28& -22.79& 2.40&    QSO&\\
J090745.28+532421.4& 0.7115& 0.36&   7/---& -----&  ---&17.60&17.37&17.54& -24.80& 1.36&    QSO&\\
J090835.85+415046.3& 0.7337& 0.04& 114/222& 229.2& 0.03&19.30&18.76&18.79& -23.42& 3.13&   QSO&2\\
J090924.66+274402.8& 0.3342& 0.07&  57/ 47& 117.6& 0.74&18.42&18.50&18.42& -22.16& 2.67&    Sy1&\\
J090924.68+521632.6& 0.4102& 0.18&  33/---& -----&  ---&18.57&17.95&17.50& -23.03& 2.22&    QSO&1,2\\
J091133.85+442250.1& 0.2976& 0.03&  30/157& 446.4& 0.84&18.76&18.72&18.24& -21.67& 2.47&    Sy1&2\\
J091925.65+110659.4& 0.4247& 0.09&  31/ 31&  34.5& 0.09&20.70&20.18&18.75& -20.88& 3.10&BL Lac?&2\\
J092414.70+030900.8& 0.1280& 2.71&  19/---& -----&  ---&18.78&18.13&17.19& -20.35& 2.01&    Sy1&2\\
J092502.41+105202.0& 0.3185& 0.20&  12/---&  28.8& 0.70&24.00&21.22&19.00& -18.85& 3.08& Galaxy&\\
J093200.08+553347.4& 0.2656& 1.27&   7/---& -----&  ---&17.40&17.44&17.30& -22.73& 1.32&    QSO&\\
J093209.60+363002.5& 0.1537& 0.05&  27/---& -----&  ---&20.05&18.87&17.47& -19.91& 2.46&BL Lac?&\\
J093430.68+030545.3& 0.2252& 0.07&  83/119& 310.3& 0.77&21.23&19.78&18.09& -19.77& 3.33& LINERs&\\
J093712.33+500852.1& 0.2756& 0.19& 212/315& 170.8&-0.49&19.65&19.10&18.07& -21.04& 3.47&    Sy1&2\\
J094857.31+002225.5& 0.5846& 0.19& 127/---& -----&  ---&18.60&18.36&18.12& -23.41& 3.00&  NLSy1&2\\
J095855.10+423704.0& 0.6641& 1.88&  23/ 60&  98.0& 0.39&19.09&18.77&18.93& -23.25& 2.43&   QSO&\\
J101244.30+422957.0& 0.3651& 0.13&  28/---& -----&  ---&18.89&18.57&17.91& -22.21& 2.39& BL Lac&1,2\\
J101258.34+393238.8& 0.1709& 0.04&  18/---& -----&  ---&19.95&19.03&17.86& -20.02& 2.35&BL Lac?&2\\
J101447.77+442133.3& 0.7955& 0.09&   5/ 22&  67.9& 0.91&19.31&18.87&18.96& -23.48& 1.83&   QSO&2\\
J102044.32+492046.2& 0.3899& 0.14&   8/---& -----&  ---&18.53&18.55&18.45& -22.43& 1.85&    QSO&\\
J102106.04+452331.8& 0.3642& 0.61&  22/ 82& 127.2& 0.35&18.23&18.19&18.35& -22.63& 2.14&   QSO&1,2\\
J102235.57+454105.4& 0.7434& 0.05&  12/ 59& 145.6& 0.73&21.06&20.81&19.96& -21.44& 2.98&    Sy1&\\
J102237.44+393150.1& 0.6036& 0.90&  40/---& -----&  ---&17.68&17.32&17.01& -24.49& 2.09&    QSO&2\\
J102504.20+414332.6& 0.6818& 0.04&   6/ 20&   8.9&-0.65&18.99&18.43&18.60& -23.59& 1.72&   QSO&2\\
J102523.04+040229.0& 0.2077& 0.10&  31/---& -----&  ---&19.54&19.01&18.09& -20.53& 2.59&BL Lac?&2\\
J102738.53+605016.5& 0.3320& 0.07&   8/---& -----&  ---&17.76&17.57&17.32& -23.03& 1.44&    QSO&\\
J103024.95+551622.7& 0.4346& 0.11&   6/ 80& 156.1& 0.54&17.18&16.94&16.68& -24.22& 1.09&   QSO&\\
J103144.75+602030.4& 1.2296& 0.89& 231/253& 320.8& 0.19&18.22&18.17&17.89& -25.06& 3.26&   QSO&2\\
J103214.53+635950.3& 0.5564& 0.16&  17/ 31&  34.3& 0.08&18.80&18.48&18.57& -23.17& 2.17&   QSO&\\
J103854.49+513937.4& 0.4701& 0.07&  21/ 22&   7.8&-0.83&23.36&22.55&20.01& -18.68& 3.88& Galaxy&\\
J104149.15+390119.5& 0.2084& 0.10&  23/ 19&  33.4& 0.45&20.01&19.24&17.97& -20.27& 2.55&BL Lac?&\\
J104334.81+343232.5& 0.7330& 0.18&  49/ 42& 113.5& 0.80&20.92&20.18&20.23& -21.96& 3.33&    Sy1&\\
J104410.67+532220.5& 1.9033& 0.10& 307/437& 493.0& 0.10&19.41&19.38&19.11& -24.59& 3.92&   QSO&2\\
J105431.89+385521.6& 1.3664& 1.89&  47/ 56&  61.2& 0.07&17.24&17.15&17.17& -26.26& 2.17&   QSO&\\
J105837.73+562811.1& 0.1433& 0.11& 177/247& 228.0&-0.06&16.84&16.42&15.95& -22.34& 2.30& BL Lac&1,2\\
J110409.63+122157.5& 0.5908& 0.15&  33/---& -----&  ---&17.78&17.30&17.48& -24.45& 1.99&    QSO&\\
J111421.76+582319.8& 0.2057& 0.09&   2/---&  65.7& 2.81&22.01&20.38&18.60& -18.95& 1.95& LINERs&3\\
J111908.94+090022.8& 0.3316& 0.30&  13/---&  22.3& 0.43&23.13&21.09&19.09& -19.19& 3.07& LINERs&2\\
J113251.05+631144.0& 0.1114& 0.26&   7/---& -----&  ---&20.50&18.42&16.98& -19.51& 1.69& Galaxy&\\
J113255.96+051539.6& 0.1009& 0.06&  78/---& -----&  ---&20.04&18.35&17.10& -19.43& 2.70& LINERs&\\
J113518.79+125311.1& 0.2040& 0.13&  64/ 57& 143.9& 0.75&20.20&19.30&18.01& -20.14& 3.02& LINERs&2\\
J114510.39+011056.2& 0.6260& 0.02&  15/---&  41.9& 0.83&19.29&18.98&18.98& -22.92& 2.32&   QSO&2\\
J114803.17+565411.4& 0.4511& 0.04&  25/ 51&  65.9& 0.21&17.93&17.70&17.75& -23.54& 2.01&   QSO&2\\
J115227.48+320959.4& 0.3746& 0.46&  45/---& -----&  ---&19.22&19.08&18.55& -21.78& 2.81&    Sy1&\\
J115232.86+493938.7& 1.0930& 0.07& 215/167& 110.3&-0.33&17.36&17.17&16.91& -25.82& 2.81&   QSO&2\\
J115323.95+583138.4& 0.2024& 0.29& 105/---& -----&  ---&20.04&19.64&18.17& -19.87& 3.37&    Sy1&2\\
J115326.70+361726.3& 1.3574& 0.09&  49/ 60&  60.5& 0.01&17.35&17.23&17.10& -26.16& 2.22&   QSO&\\
J115409.28+023815.1& 0.2105& 0.26&  59/ 87& 115.2& 0.23&20.81&19.67&18.13& -19.80& 3.13&    Sy1&2\\
J115700.59+324457.8& 0.4862& 0.12&  86/ 82& 159.9& 0.54&18.54&18.33&18.39& -23.07& 2.80&   QSO&\\
J115727.60+431806.3& 0.2300& 0.16&  93/106& 256.0& 0.71&18.34&18.30&17.55& -21.54& 2.79&    Sy1&\\
J120303.50+603119.1& 0.0653& 0.16& 145/182& 190.3& 0.04&17.62&16.89&15.98& -20.11& 2.38& BL Lac&2\\
J120335.39+451049.5& 1.0760& 0.16&  50/ 43&  38.5&-0.09&17.89&17.65&17.38& -25.30& 2.37&   QSO&1,2\\
J120436.17+485653.9& 0.4509& 2.10&  39/---& -----&  ---&17.83&17.72&17.83& -23.54& 2.21&    QSO&2\\
J120837.11+612106.4& 0.2748& 0.14&  56/118& 180.3& 0.34&19.50&18.86&17.65& -21.26& 2.80& BL Lac&\\
J121509.95+462715.1& 0.7201& 1.58& 178/168& 271.5& 0.39&17.69&17.34&17.58& -24.83& 2.76&   QSO&\\
J122313.21+540906.5& 0.1558& 0.14&  41/140& 387.6& 0.82&17.51&17.34&16.68& -21.64& 2.03&    Sy1&1,2\\
J122353.05+465048.8& 0.2606& 0.10&   9/---&  12.8& 0.28&20.43&19.57&18.20& -20.40& 2.29&BL Lac?&2\\
J122424.23+401510.5& 0.4169& 0.08&  29/---& -----&  ---&18.05&17.89&18.02& -23.20& 2.15&    QSO&\\
J122506.50+483435.1& 0.6467& 1.17&  19/ 38&  76.1& 0.56&19.72&19.25&19.31& -22.68& 2.54&   QSO&2\\
J122751.22+632305.3& 0.1455& 0.00&   5/ 71& 242.0& 0.99&21.03&18.98&17.51& -19.54& 1.77& Galaxy&\\
J123157.08+542028.7& 0.5158& 0.26&  13/ 26&  61.0& 0.69&19.35&19.10&19.03& -22.41& 2.30&   QSO&2\\
J123807.76+532556.0& 0.7640& 0.10&  37/ 56& 114.6& 0.58&17.36&17.32&17.36& -25.02& 2.07&   QSO&2\\
J123819.62+412420.5& 0.4980& 0.04&   9/---& -----&  ---&18.59&18.29&18.33& -23.14& 1.81&    QSO&\\
J124139.72+493405.5& 0.4737& 0.07&   6/ 40& 140.2& 1.01&17.72&17.38&17.36& -23.94& 1.27&   QSO&\\
J124834.30+512807.8& 0.3509& 0.15&  55/ 88& 114.1& 0.21&18.65&18.19&17.39& -22.48& 2.53& BL Lac&1,2,c\\
J125223.78+645137.9& 0.3120& 0.21&  20/ 25&  42.3& 0.42&18.01&18.24&18.21& -22.30& 2.11&   QSO&\\
J125303.17+450044.8& 0.0777& 0.25&  11/---& -----&  ---&19.87&18.12&16.68& -19.09& 1.76& Galaxy&\\
J125326.15+505428.3& 0.1217& 0.11&   4/---& -----&  ---&20.78&18.81&17.48& -19.33& 1.60& Galaxy&1\\
J130612.15+514407.0& 0.2773& 0.07&  15/---& -----&  ---&22.83&20.62&18.68& -19.25& 2.93& Galaxy&\\
J131211.14+480925.3& 0.7151& 0.15&  58/---& -----&  ---&17.50&17.12&17.28& -25.03& 2.18&    QSO&2\\
J131931.73+140533.1& 0.5729& 0.14&  52/ 57&  76.2& 0.23&18.14&17.93&17.40& -23.81& 2.44& BL Lac&\\
J132419.67+041907.0& 0.2631& 0.22&  13/ 46& 161.7& 1.01&22.49&20.62&18.94& -19.20& 2.86& LINERs&2\\
J132631.44+473755.8& 0.6821& 0.02&  12/---& -----&  ---&19.00&18.60&18.65& -23.45& 2.08&    QSO&\\
J133245.24+472222.6& 0.6693& 0.12& 302/333& 233.2&-0.29&18.09&17.66&17.10& -24.35& 3.11&   QSO&2\\
J133437.49+563147.9& 0.3428& 1.21&  36/107& 185.5& 0.44&18.35&18.40&18.50& -22.31& 2.43&   QSO&1,2\\
J133655.51+654116.0& 0.4375& 0.50&   6/ 69& 206.4& 0.88&18.60&18.35&18.17& -22.82& 1.65&   QSO&1,2\\
J134136.23+551437.9& 0.2069& 0.13&  30/ 34&  37.1& 0.07&19.35&18.79&17.83& -20.74& 2.49& BL Lac&2\\
J134545.35+533252.2& 0.1353& 0.31&  70/223& 429.8& 0.53&18.35&18.08&17.09& -20.58& 2.56&    Sy1&2\\
J134617.54+622045.4& 0.1164& 0.30&   7/---& -----&  ---&17.15&17.08&16.27& -21.29& 1.15&    Sy1&1,2\\
J134751.58+283629.7& 0.7399& 1.27&  41/ 92& 290.0& 0.92&17.95&17.44&17.54& -24.75& 2.16&   QSO&\\
J135229.02+490823.0& 0.3966& 0.20&  61/124&  70.4&-0.46&19.17&19.11&19.01& -21.88& 2.96&    Sy1&\\
J135305.54+044338.6& 0.5234& 0.81&  58/ 95& 154.5& 0.39&17.99&17.75&17.91& -23.79& 2.41&   QSO&2\\
J135341.72+431052.5& 1.1140& 0.16&  15/---& -----&  ---&17.44&17.24&17.10& -25.79& 1.69&    QSO&1,2\\
J141149.43+524900.1& 0.0765& 0.25&  96/354& 846.8& 0.70&18.58&17.33&16.14& -19.93& 2.38& Galaxy&2\\
J141159.73+423950.3& 0.8861& 0.13&  46/---& -----&  ---&17.19&17.02&17.00& -25.58& 2.06&    QSO&1,2\\
J141446.64+392818.6& 0.6573& 0.14&  11/---& -----&  ---&19.91&19.46&19.41& -22.51& 2.39&    QSO&2\\
J141628.66+124213.5& 0.3346& 0.05&  75/ 98& 110.2& 0.09&17.41&17.44&17.65& -23.21& 2.37&   QSO&2\\
J141740.44+381821.1& 0.4495& 0.06& 173/ 95& 111.3& 0.13&18.45&18.32&18.31& -22.93& 3.10&   QSO&2\\
J142020.67+462440.8& 1.2445& 0.17&  30/ 34&  46.3& 0.25&19.61&19.77&19.23& -23.52& 3.01&   QSO&2\\
J142106.03+385522.7& 0.4888& 0.30&  97/103&  85.7&-0.15&18.89&18.59&18.45& -22.80& 2.96&   QSO&2\\
J142314.19+505537.3& 0.2759& 2.61& 125/---& -----&  ---&17.75&17.93&17.71& -22.34& 2.77&    QSO&2\\
J142421.17+370552.8& 0.2896& 1.79&  11/ 29&  78.6& 0.80&20.41&19.65&18.20& -20.56& 2.41&BL Lac?&2\\
J142606.19+402432.0& 0.6639& 0.29&  18/---& -----&  ---&20.41&19.91&19.63& -22.08& 2.78&    Sy1&\\
J142730.27+540923.7& 0.1060& 0.80&  23/ 26&  44.0& 0.42&19.76&18.46&17.15& -19.49& 2.22& Galaxy&1\\
J143726.14+504555.8& 0.7835& 0.56&   6/---& -----&  ---&17.86&17.53&17.57& -24.81& 1.37&    QSO&1\\
J143942.83+582759.2& 0.4249& 0.02&  13/ 45&  46.3& 0.02&18.02&17.91&18.09& -23.22& 1.81&   QSO&1,2\\
J144542.78+390921.4& 0.1625& 0.81&   7/---& -----&  ---&22.19&20.64&19.29& -18.20& 2.59&    Sy2&\\
J145224.68+452223.6& 0.4676& 2.10& 123/---& -----&  ---&16.97&16.80&16.87& -24.52& 2.35&    QSO&\\
J145247.37+473529.1& 1.1576& 0.04&  14/---& -----&  ---&18.94&18.85&18.64& -24.26& 2.31&    QSO&\\
J145958.43+333701.8& 0.6449& 0.15& 103/ 48&  15.3&-0.92&16.88&16.62&16.77& -25.35& 2.22&   QSO&1,2\\
J150117.97+545518.2& 0.3386& 0.15&  18/---& -----&  ---&20.56&19.92&18.15& -20.64& 2.74&    Sy2&2\\
J150324.77+475829.6& 0.3445& 0.08&   9/ 56& 111.2& 0.55&18.39&18.00&17.30& -22.65& 1.67& LINERs&\\
J150455.56+564920.2& 0.3590& 1.39&   5/---& -----&  ---&17.06&17.00&17.09& -23.79& 1.02&    QSO&1\\
J151017.82+422155.0& 0.4876& 0.04&  19/111& 262.5& 0.69&21.60&21.20&19.85& -20.17& 3.30&    Sy1&2\\
J151830.93+483214.4& 0.5757& 0.16&   4/---& -----&  ---&18.66&18.27&18.32& -23.44& 1.46&    QSO&1\\
J151838.90+404500.2& 0.0652& 0.07&  25/---& -----&  ---&19.28&17.63&16.56& -19.21& 1.91& LINERs&1\\
J151844.76+461855.1& 0.8852& 0.13&  35/163& 436.0& 0.79&20.20&19.91&19.80& -22.67& 3.10&   QSO&\\
J151913.35+362343.4& 0.2857& 0.11&  59/---& -----&  ---&19.49&19.51&18.79& -20.81& 3.08&    Sy1&\\
J152556.22+591659.5& 0.9551& 0.08&   7/---& -----&  ---&18.51&18.33&18.21& -24.42& 1.77&    QSO&\\
J153102.48+435637.6& 0.4520& 0.10&  18/---& -----&  ---&17.31&17.13&16.91& -24.12& 1.64&    QSO&1,2\\
J153253.78+302059.3& 0.3622& 0.14&   8/---& -----&  ---&20.65&20.00&19.15& -20.71& 2.42& LINERs&2,3\\
J153447.20+371554.5& 0.1428& 0.86&  18/ 24&  22.0&-0.07&18.28&17.77&17.00& -20.97& 1.84& BL Lac&2\\
J154232.03+493842.5& 0.5897& 0.12&  24/ 34&  53.8& 0.37&19.58&19.16&19.26& -22.60& 2.60&   QSO&2\\
J154535.56+532421.1& 0.2082& 0.10&   4/ 49& 216.8& 1.20&22.06&19.89&18.11& -19.38& 2.05& Galaxy&\\
J160051.34+331205.9& 0.2835& 0.69&   4/ 32& 108.4& 0.98&24.12&20.92&18.87& -18.83& 2.48& Galaxy&\\
J160239.61+264606.0& 0.3716& 0.19&  98/---& -----&  ---&22.35&21.29&19.82& -19.40& 4.03& LINERs&\\
J160317.91+090037.9& 0.4884& 2.93&   6/ 99& 203.7& 0.58&18.35&17.98&17.98& -23.40& 1.51&   QSO&\\
J160658.30+271705.5& 0.9335& 0.17& 143/228& 178.0&-0.20&19.23&18.88&18.79& -23.79& 3.30&   QSO&\\
J160813.79+292126.3& 1.2006& 1.08&  12/---&  38.9& 0.95&18.87&19.08&18.67& -24.16& 2.33&   QSO&\\
J160822.16+401217.9& 0.6282& 0.12& 121/240& 200.9&-0.14&21.09&20.68&19.89& -21.21& 3.91&    Sy1&2\\
J161706.32+410647.0& 0.2667& 0.03&  81/124&  95.1&-0.21&19.03&18.48&17.57& -21.59& 2.81& BL Lac&\\
J161826.93+081950.7& 0.4459& 0.33& 150/119& 130.7& 0.08&17.30&17.02&16.78& -24.19& 2.52&   QSO&\\
J161902.49+303051.5& 1.2878& 0.21&  37/---& -----&  ---&16.95&16.84&16.52& -26.46& 1.94&    QSO&1,2\\
J162111.27+374604.9& 1.2734& 0.09& 154/201& 631.0& 0.92&21.22&19.85&18.27& -23.22& 3.76&   QSO&2\\
J162229.31+400643.5& 0.6877& 1.15&  27/ 58&  36.9&-0.36&18.81&18.48&18.63& -23.60& 2.39&   QSO&1,2\\
J162358.25+074130.5& 1.2970& 0.55& 170/149& 100.1&-0.32&17.89&18.01&17.59& -25.35& 3.07&   QSO&\\
J162711.89+314359.3& 0.7324& 0.33&  36/ 76& 127.9& 0.42&20.92&20.34&20.10& -21.83& 3.26&    Sy1&\\
J163624.31+471535.9& 0.8232& 0.08&   5/---&  19.1& 1.08&19.12&18.79&18.93& -23.64& 1.80&   QSO&\\
J163726.66+454748.9& 0.1922& 0.12&  11/ 20&  45.3& 0.66&19.78&19.01&17.83& -20.32& 2.14&BL Lac?&\\
J163856.53+433512.5& 0.3392& 1.21&   5/ 50& 191.8& 1.08&18.48&18.45&18.27& -22.22& 1.60&    Sy1&2\\
J164054.16+314329.9& 0.9571& 0.87&  27/ 41&  60.2& 0.31&20.39&20.03&19.93& -22.68& 3.04&   QSO&\\
J164442.53+261913.2& 0.1442& 0.13&  92/ 99& 131.6& 0.23&18.03&18.02&17.43& -20.82& 2.66&  NLSy1&2\\
J165005.47+414032.4& 0.5850& 0.38& 100/136& 232.0& 0.43&17.88&17.55&17.52& -24.20& 2.57&   QSO&2\\
J170112.38+353353.4& 0.5013& 0.08& 130/ 50&  70.7& 0.28&18.55&18.33&18.39& -23.13& 2.99&   QSO&2\\
J170123.97+385136.9& 1.1132& 0.13&  49/ 61& 198.5& 0.95&18.84&18.95&18.66& -24.13& 2.89&   QSO&2\\
J171322.58+325627.9& 0.1014& 0.22&  25/---& -----&  ---&18.00&17.77&16.91& -20.27& 1.98&    Sy1&1\\
J172010.03+263732.0& 0.1592& 0.14&   7/---& -----&  ---&20.78&19.32&18.11& -19.49& 2.06& LINERs&3\\
J172242.16+281500.0& 0.9468& 0.12& 150/224& 249.4& 0.09&18.20&18.03&18.01& -24.70& 2.98&   QSO&1,2\\
J205456.85+001537.7& 0.1508& 0.32&  38/ 65&  56.5&-0.11&19.07&18.45&17.52& -20.38& 2.44&BL Lac?&2\\
\hline
\end{longtable}
\end{center}
\tablecomments{0.53\textwidth}{
 (1): SDSS name --- Jhhmmss.s+ddmmss.s\\
 (2): z --- redshift\\
 (3): ${\Delta}_{ro}$ --- the distance between the position of radio and optical in units of arcsec\\
 (4): $F_{6cm}$  --- flux of 4.85GHz core and total in units of mJy\\
 (5): $F_{20cm}$ --- flux of NVSS 1.40GHz in units of mJy\\
 (6): $\alpha_{r}$ --- radio spectral index between 4.85GHz and 1.40GHz\\
\[
{\alpha}_{r}=-\frac{log(F_{4.85GHz}/F_{1.4GHz})}{log({\nu}_{4.85GHz}/{\nu}_{1.4GHz})}
\]
 (7): $u$ --- Galactic extinction corrected psf magnitude of SDSS u band\\
 (8): $g$ --- Galactic extinction corrected psf magnitude of SDSS g band\\
 (9): $i$ --- Galactic extinction corrected psf magnitude of SDSS i band\\
(10): $M_{B}$ --- Galactic extinction corrected B band absolute magnitude\\
(11): $\lg R$ --- log radio loudness, radio loudness follow \cite{kel89}\\
(12): SpecClass --- SDSS spectral class\\
(13): ref --- reference about source spectral identification\\
              No reference suggests that it is firstly identified by our work.\\
              1:Laurent-Muehleisen et~al.~\cite{LM98} \\
              2:Turriziani et~al.~\cite{Tur07}\\
              3:Crawford et~al.~\cite{craw}\\
              }
\end{landscape}

%\begin{landscape}
\begin{center}
\begin{longtable}{crrrrrrrr}
\caption{SDSS Narrow Emission Line Parameters}\\
\label{tab:nlelp}\\
\hline
SDSS name&F($H\beta$)&F([OIII]&F($H\alpha$)&F([NII])&F([SII])&F([OI])&F([OII])&SpecClass\\
         &$\lambda 4861$&$\lambda 5007$&$\lambda 6563$&$\lambda 6583$&$\lambda 6716$/$\lambda 6731$&$\lambda 6300$&$\lambda 3727$&\\
(1)&(2)&(3)&(4)&(5)&(6)&(7)&(8)&(9)\\
\hline
\endfirsthead
\hline
SDSS name&F($H\beta$)&F([OIII]&F($H\alpha$)&F([NII])&F([SII])&F([OI])&F([OII])&SpecClass\\
         &$\lambda 4861$&$\lambda 5007$&$\lambda 6563$&$\lambda 6583$&$\lambda 6716$/$\lambda 6731$&$\lambda 6300$&$\lambda 3727$&\\
(1)&(2)&(3)&(4)&(5)&(6)&(7)&(8)&(9)\\
\hline
\endhead
\hline
J074906.50+451033.9& 308&3557&1600&1026& 213/ 143&  86& 562&Sy1\\
J075244.19+455657.3&  54& 899& 282& 643& 205/ 202& 233& 362&Sy1\\
J080131.96+473616.0& 163&1292& 228&1251& 120/   0&  31&   0&Sy1\\
J084203.73+401831.3&  75& 595& 222& 115&  44/  57&  23&  29&Sy1\\
J085004.65+403607.7&  14&  64&  66& 124&  26/  25&  37&  50&LINER\\
J085348.18+065447.1&  17& 354&  88& 108&  38/  38&  49&  82&Sy1\\
J091133.85+442250.1&  78& 325& 421& 229&  46/  18&   7& 102&Sy1\\
J092414.70+030900.8&  87& 624& 420& 315&  76/  72& 104& 147&Sy1\\
J093200.08+553347.4&  99& 399& 439& 206&  55/  51&  13&  44&Sy1\\
J093430.68+030545.3&  35& 108& 180& 213&  97/ 112&  42& 195&LINER\\
J111421.76+582319.8&  25&  23&  93& 117&  48/  35&  30& 104&LINER\\
J111908.94+090022.8&   9&  17&  47&  75&  16/   0&   9&  20&LINER\\
J113255.96+051539.6&  33& 261&  71& 154&  49/  58&  39& 100&LINER\\
J113518.79+125311.1&  22&  43&  66& 106&  48/  40&  24&  95&LINER\\
J115323.95+583138.4&  75& 630& 203&  99&  51/  50& 101& 168&Sy1\\
J115409.28+023815.1&  19& 155&  91&  73&  29/  28&  33& 130&Sy1\\
J115727.60+431806.3& 134&1417& 402& 207&  88/  79&  41& 287&Sy1\\
J122313.21+540906.5& 352&2595&1490& 822& 319/ 266& 106& 782&Sy1\\
J132419.67+041907.0&  32&   7&  99&  99&  49/  39&  15&  99&LINER\\
J133437.49+563147.9&  12& 316&  67&  90&  23/  21&   8&  48&Sy1\\
J134545.35+533252.2&  27& 342&  90& 130&  61/  55&  52&  87&Sy1\\
J134617.54+622045.4& 121&1332& 426& 213& 106/  93&  72& 583&Sy1\\
J144542.78+390921.4&  13& 146&  83&  66&  25/  18&  10&  31&Sy2\\
J150117.97+545518.2&  56& 779& 218& 326&  55/  67&  45& 173&Sy2\\
J150324.77+475829.6&  54& 180& 140& 162&  40/  68&  12&   0&LINER\\
J151838.90+404500.2&  50& 185& 252& 352& 156/ 130&  71& 292&LINER\\
J151913.35+362343.4&  43& 300& 159& 104&  34/  40&  34& 111&Sy1\\
J153253.78+302059.3& 248& 145& 880& 653&   0/   0& 216& 855&LINER\\
J160239.61+264606.0&  57& 119& 277& 439&   0/   0&  91& 273&LINER\\
J164442.53+261913.2& 331& 164&1065& 242&   0/  13&   0&  31&Sy1\\
J171322.58+325627.9& 107&1069& 521& 460& 161/ 134&  57& 261&Sy1\\
J172010.03+263732.0&  91&  38& 324& 281& 128/ 102&  87& 272&LINER\\
\hline
\end{longtable}
%1\tablenotetext{
\tablecomments{0.53\textwidth}{
(1): SDSS name --- Jhhmmss.s+ddmmss.s\\
% (2): F($H\beta$)  --- $H\beta$ narrow emission line flux in Unit:${10}^{-17}\,erg\,s^{-1}\,{cm}^{-2}$\\
% (3): F([OIII] --- [OIII] narrow emission line flux in Unit:${10}^{-17}\,erg\,s^{-1}\,{cm}^{-2}$\\
% (4): F($H\alpha$)  --- $H\alpha$ narrow emission line flux in Unit:${10}^{-17}\,erg\,s^{-1}\,{cm}^{-2}$\\
% (5): F([NII]) --- [NII] narrow emission line flux in Unit:${10}^{-17}\,erg\,s^{-1}\,{cm}^{-2}$\\
% (6): F([SII]) --- [SII] narrow emission line flux in Unit:${10}^{-17}\,erg\,s^{-1}\,{cm}^{-2}$\\
% (7): F([OI])  --- [OI] narrow emission line flux in Unit:${10}^{-17}\,erg\,s^{-1}\,{cm}^{-2}$\\
% (8): F([OII]) --- [OII] narrow emission line flux in Unit:${10}^{-17}\,erg\,s^{-1}\,{cm}^{-2}$\\
(2)-(8):Narrow emission line flux in units of ${10}^{-17}\,erg\,s^{-1}\,{cm}^{-2}$\\
(9): SpecClass --- SDSS spectral class\\
 }
%}
%\end{landscape}

\begin{longtable}{cccrrccc}
\caption{SDSS Broad Emission Line Parameters}\\
\label{bel}\\
\hline
SDSS name&z&line&flux&FWHM&$lg L_{BLR}$&$lg M_{BH}$&SpecClass\\
(1)&(2)&(3)&(4)&(5)&(6)&(7)&(8)\\
\hline
\endfirsthead
\hline
SDSS name&z&line&flux&FWHM&$lg L_{BLR}$&$lg M_{BH}$&SpecClass\\
(1)&(2)&(3)&(4)&(5)&(6)&(7)&(8)\\
\hline
\endhead
\hline
J000132.83+145608.0& 0.3989& $H\beta$&   535&  3175&43.88& 8.0&    Sy1\\
J011354.50+132452.4& 0.6854& $H\beta$&   563&  3699&44.47& 8.5&   QSO\\
J073320.84+390505.2& 0.6638& $H\beta$&   496&  3471&44.38& 7.5&   QSO\\
J074906.50+451033.9& 0.1922&$H\alpha$&  5462&  3779&43.62& 8.3&    QSO\\
                   &       & $H\beta$&  7535&  4840&44.30& 8.6&       \\
J075244.19+455657.3& 0.0518&$H\alpha$&  4058&  3067&42.27& 7.5&    Sy1\\
J075448.86+303355.1& 0.7955& $H\beta$&  1798&  8502&45.13& 8.6&    QSO\\
J080131.96+473616.0& 0.1569& $H\beta$& 11568&  7287&44.29& 9.0&    QSO\\
                   &       &$H\alpha$&  9426&  6319&43.66& 8.7&       \\
J080644.42+484149.2& 0.3700& $H\beta$&  1076& 14698&44.11& 9.5&    QSO\\
J081058.99+413402.7& 0.5067& $H\beta$&   842&  3483&44.32& 8.3&   QSO\\
J081100.60+571412.5& 0.6110& $H\beta$&  2226&  4568&44.94& 8.0&   QSO\\
J081432.11+560956.6& 0.5094& $H\beta$&   782&  2300&44.30& 8.0&   QSO\\
J081902.32+322637.2& 0.6512& $H\beta$&   139& 17188&43.81& 9.4&    Sy1\\
J081916.61+264203.2& 0.5261& $H\beta$&   368&  6880&44.00& 8.7&   QSO\\
J084203.73+401831.3& 0.1516& $H\beta$&  4188&  8810&43.82& 8.8&    Sy1\\
                   &       &$H\alpha$&  3358&  7774&43.18& 8.7&       \\
J084650.00+064149.0& 0.6161& $H\beta$&  2751&  7836&45.04& 9.5&   QSO\\
J085039.95+543753.3& 0.3673& $H\beta$&  1079& 12595&44.10& 9.3&    Sy1\\
J085348.18+065447.1& 0.2234&$H\alpha$&   704&  3813&42.87& 7.9&    Sy1\\
J085830.61+080422.8& 0.4549& $H\beta$&  1132&  6634&44.34& 8.9&    QSO\\
J090745.28+532421.4& 0.7113& $H\beta$&  1074&  7213&44.79& 9.3&    QSO\\
J090835.85+415046.3& 0.7337& $H\beta$&   867&  3545&44.73& 8.6&   QSO\\
J090924.66+274402.8& 0.3342& $H\beta$&   454&  9815&43.63& 8.8&    Sy1\\
J090924.68+521632.6& 0.4102& $H\beta$&  1507& 10542&44.36& 9.3&    QSO\\
J091133.85+442250.1& 0.2976& $H\beta$&   873&  2861&43.80& 7.8&    Sy1\\
                   &       &$H\alpha$&   734&  2688&43.18& 7.8&       \\
J092414.70+030900.8& 0.1280&$H\alpha$&   460&  6140&42.16& 8.0&    Sy1\\
                   &       & $H\beta$&  2087&  6148&43.36& 7.5&       \\
J093200.08+553347.4& 0.2657& $H\beta$&  2104&  4531&44.06& 8.4&    QSO\\
                   &       &$H\alpha$&  1738&  4198&43.44& 8.3&       \\
J093712.33+500852.1& 0.2756&$H\alpha$&   123&  2922&42.32& 7.5&    Sy1\\
                   &       & $H\beta$&   543&  2923&43.51& 6.9&       \\
J094857.31+002225.5& 0.5846& $H\beta$&   301&  1850&44.03& 7.6&  NLSy1\\
J095855.10+423704.0& 0.6641& $H\beta$&   415&  2817&44.30& 7.3&   QSO\\
J101447.77+442133.3& 0.7956& $H\beta$&   320&  5468&44.38& 8.8&   QSO\\
J102044.32+492046.2& 0.3899& $H\beta$&   898&  4396&44.08& 7.5&  QSO\\
J102106.04+452331.8& 0.3644& $H\beta$&  1086&  7669&44.09& 8.9&   QSO\\
J102235.57+454105.4& 0.7432& $H\beta$&   424& 16214&44.43& 9.7&    Sy1\\
J102237.44+393150.1& 0.6036& $H\beta$&   473&  4747&44.26& 8.6&    QSO\\
J102504.20+414332.6& 0.6818& $H\beta$&   706&  6159&44.56& 9.0&   QSO\\
J102738.53+605016.5& 0.3320& $H\beta$&  1312& 23649&44.08& 9.9&    QSO\\
                   &       &$H\alpha$&   727& 15006&43.28& 9.4&       \\
J103024.95+551622.7& 0.4345& $H\beta$&  3828&  2244&44.82& 8.3&   QSO\\
J103144.75+602030.4& 1.2303&     MgII&   465&  3041&44.82& ---&   QSO\\
J103214.53+635950.3& 0.5562& $H\beta$&   536&  5623&44.23& 8.7&   QSO\\
J104334.81+343232.5& 0.7330& $H\beta$&   197&  6264&44.08& 8.7&    Sy1\\
J104410.67+532220.5& 1.9008&     MgII&   236&  2591&44.99& ---&   QSO\\
J105431.89+385521.6& 1.3661&     MgII&   352&  2452&44.81& ---&   QSO\\
J110409.63+122157.5& 0.5907& $H\beta$&  1791&  5493&44.81& 8.1&    QSO\\
J114510.39+011056.2& 0.6261& $H\beta$&   264&  6572&44.04& 8.7&   QSO\\
J114803.17+565411.4& 0.4510& $H\beta$&  1979&  6159&44.57& 9.0&   QSO\\
J115227.48+320959.4& 0.3747& $H\beta$&   527&  5080&43.81& 8.3&    Sy1\\
J115232.86+493938.7& 1.0929&     MgII&  1991&  3801&45.33& ---&   QSO\\
J115323.95+583138.4& 0.2024&$H\alpha$&  1203&  6066&43.01& 8.4&    Sy1\\
                   &       & $H\beta$&  4592&  6066&44.14& 7.8&       \\
J115326.70+361726.3& 1.3580&     MgII&   371&  2034&44.83& ---&   QSO\\
J115409.28+023815.1& 0.2106&$H\alpha$&   602&  3278&42.75& 7.8&    Sy1\\
J115700.59+324457.8& 0.4862& $H\beta$&   898&  2318&44.31& 8.0&   QSO\\
J115727.60+431806.3& 0.2300&$H\alpha$&   670&  2867&42.88& 7.7&    Sy1\\
                   &       & $H\beta$&   952&  3274&43.58& 7.8&       \\
J120335.39+451049.5& 1.0766&     MgII&  1230&  3726&45.10& ---&   QSO\\
J120436.17+485653.9& 0.4509& $H\beta$&   965&  3422&44.26& 8.3&    QSO\\
J121509.95+462715.1& 0.7201& $H\beta$&  2276&  4550&45.13& 9.1&   QSO\\
J122313.21+540906.5& 0.1559& $H\beta$&  3118&  6708&43.72& 8.5&    Sy1\\
                   &       &$H\alpha$&  2221&  5111&43.03& 8.3&       \\
J122424.23+401510.5& 0.4164& $H\beta$&  1879&  5500&44.47& 8.8&    QSO\\
J122506.50+483435.1& 0.6468& $H\beta$&   534&  5092&44.38& 8.7&   QSO\\
J123157.08+542028.7& 0.5156& $H\beta$&   337&  6794&43.94& 7.9&   QSO\\
J123807.76+532556.0& 0.3475&$H\alpha$&  2021& 14772&43.77& 9.6&   QSO\\
                   &       & $H\beta$&  2206& 13890&44.35& 9.6&       \\
J123819.62+412420.5& 0.4980& $H\beta$&  1041&  9322&44.40& 9.2&    QSO\\
J124139.72+493405.5& 0.4739& $H\beta$&  1785&  8082&44.58& 9.2&   QSO\\
J125223.78+645137.9& 0.3119& $H\beta$&  1579&  4217&44.10& 8.4&   QSO\\
                   &       &$H\alpha$&  1330&  4322&43.48& 8.3&       \\
J131211.14+480925.3& 0.7149& $H\beta$&  2411&  4673&45.14& 8.1&    QSO\\
J132631.44+473755.8& 0.6822& $H\beta$&   838&  5025&44.63& 8.9&    QSO\\
J133245.24+472222.6& 0.6694& $H\beta$&   499&  3372&44.39& 8.4&   QSO\\
J133437.49+563147.9& 0.3428& $H\beta$&   784&  3662&43.89& 8.1&   QSO\\
                   &       &$H\alpha$&   633&  3397&43.25& 8.0&       \\
J133655.51+654116.0& 0.4368& $H\beta$&  1196& 15462&44.32& 9.6&   QSO\\
J134545.35+533252.2& 0.1354& $H\beta$&  2135&  5413&43.42& 8.2&    Sy1\\
                   &       &$H\alpha$&  2121&  5253&42.87& 8.2&       \\
J134617.54+622045.4& 0.1165&$H\alpha$&  2234&  5721&42.75& 8.3&    Sy1\\
                   &       & $H\beta$&  8607&  5724&43.88& 7.7&       \\
J134751.58+283629.7& 0.7399& $H\beta$&  1524&  5358&44.98& 9.1&   QSO\\
J135229.02+490823.0& 0.3966& $H\beta$&   441&  4704&43.79& 8.3&    Sy1\\
J135305.54+044338.6& 0.5234& $H\beta$&  1098& 10857&44.47& 9.4&   QSO\\
J135341.72+431052.5& 1.1136&     MgII&   494&  2291&44.74& ---&    QSO\\
J141159.73+423950.3& 0.8865&     MgII&  1838&  4069&45.07& ---&    QSO\\
J141446.64+392818.6& 0.6571& $H\beta$&   366&  4044&44.24& 8.4&    QSO\\
J141628.66+124213.5& 0.3345& $H\beta$&  3353&  9439&44.50& 9.3&   QSO\\
J141740.44+381821.1& 0.4494& $H\beta$&   804&  2947&44.18& 8.1&   QSO\\
J142020.67+462440.8& 1.2450&     MgII&   327&  9294&44.68& ---&   QSO\\
J142106.03+385522.7& 0.4888&     MgII&   954&  3426&44.15& ---&   QSO\\
J142314.19+505537.3& 0.2759& $H\beta$&  1667&  1270&44.00& 7.3&    QSO\\
                   &       &$H\alpha$&  1352&  2848&43.37& 7.9&       \\
J142606.19+402432.0& 0.6639& $H\beta$&   421&  4168&44.31& 8.5&    Sy1\\
J143726.14+504555.8& 0.7833& $H\beta$&  2439&  5826&45.25& 9.4&    QSO\\
J143942.83+582759.2& 0.4250& $H\beta$&   969&  4575&44.20& 8.5&   QSO\\
J145224.68+452223.6& 0.4671& $H\beta$&  4419&  6566&44.96& 9.3&    QSO\\
J145247.37+473529.1& 1.1582&     MgII&   643&  6007&44.90& ---&    QSO\\
J145958.43+333701.8& 0.6448& $H\beta$&  3185&  4704&45.16& 8.1&    QSO\\
J150455.56+564920.2& 0.3589& $H\beta$&  3085&  6036&44.53& 8.9&    QSO\\
J151017.82+422155.0& 0.4873& $H\beta$&   724&  1880&44.22& 7.7&    Sy1\\
                   &       &     MgII&   205&  5621&43.48& ---&       \\
J151830.93+483214.4& 0.5757& $H\beta$&   951&  8909&44.51& 9.3&    QSO\\
J151844.76+461855.1& 0.8853&     MgII&   449&  3651&44.45& ---&   QSO\\
J151913.35+362343.4& 0.2857&$H\alpha$&   404&  8218&42.88& 8.6&    Sy1\\
                   &       & $H\beta$&  1642&  8222&44.03& 8.1&       \\
J152556.22+591659.5& 0.9551&     MgII&  1421&  8940&45.03& ---&    QSO\\
J153102.48+435637.6& 0.4520& $H\beta$&   908&  2330&44.24& 7.9&    QSO\\
J154232.03+493842.5& 0.5897& $H\beta$&   381&  5499&44.14& 7.8&   QSO\\
J160317.91+090037.9& 0.4883& $H\beta$&  1497&  9389&44.54& 9.3&   QSO\\
J160658.30+271705.5& 0.9337&     MgII&   669&  3041&44.68& ---&   QSO\\
J160813.79+292126.3& 1.2005&     MgII&   362&  6339&44.69& ---&   QSO\\
J160822.16+401217.9& 0.6281& $H\beta$&   216&  3274&43.96& 8.1&    Sy1\\
J161826.93+081950.7& 0.4455& $H\beta$&  4067& 19802&44.87&10.2&   QSO\\
J161902.49+303051.5& 1.2875&     MgII&  1209&  3351&45.29& ---&    QSO\\
J162111.27+374604.9& 1.2728&     MgII&   175&  3159&44.43& ---&   QSO\\
J162229.31+400643.5& 0.6878& $H\beta$&   814&  3545&44.63& 8.5&   QSO\\
J162358.25+074130.5& 1.2973&     MgII&   648&  3137&45.02& ---&   QSO\\
J162711.89+314359.3& 0.7324& $H\beta$&   490& 13816&44.48& 9.6&    Sy1\\
J163624.31+471535.9& 0.8233& $H\beta$&   441&  3570&44.56& 7.6&   QSO\\
J163856.53+433512.5& 0.3390& $H\beta$&  1163& 11516&44.05& 9.2&    Sy1\\
                   &       &$H\alpha$&   978& 10092&43.43& 9.1&       \\
J164054.16+314329.9& 0.9580&     MgII&   602&  5985&44.66& ---&   QSO\\
J164442.53+261913.2& 0.1443&$H\alpha$&   980&  1535&42.60& 7.0&  NLSy1\\
                   &       & $H\beta$&  2771&  1544&43.59& 6.4&       \\
J165005.47+414032.4& 0.5848& $H\beta$&  1296&  5240&44.66& 8.9&   QSO\\
J170112.38+353353.4& 0.5011& $H\beta$&   495&  2256&44.08& 7.8&   QSO\\
J170123.97+385136.9& 1.1125&     MgII&   435&  4625&44.68& ---&   QSO\\
J171322.58+325627.9& 0.1013&$H\alpha$&  1033&  3933&42.29& 7.7&    Sy1\\
                   &       & $H\beta$&  3786&  3933&43.40& 7.1&       \\
J172242.16+281500.0& 0.9447&     MgII&   663&  2762&44.69& ---&   QSO\\
\hline
\end{longtable}
%l\tablenotetext{
\tablecomments{0.53\textwidth}{
 (1): SDSS name --- Jhhmmss.s+ddmmss.s\\
 (2): z --- redshift of emission line\\
 (3): line --- broad emission line\\
 (4): flux --- broad emission line flux in units of ${10}^{-17} erg\,s^{-1}\,{cm}^{-2}$\\
 (5): FWHM --- FWHM for broad emission line in units of $km\,s^{-1}$\\
 (6): $lg L_{BLR}$ ---log broad emission line region luminosity in units of $erg\,s^{-1}$\\
 (7): $lg M_{BH}$ --- log Black Hole mass in units of Solar mass\\
 (8): SpecClass --- SDSS spectral class\\
 }
%}

\begin{longtable}{cccc}
\caption{Narrow line region luminosities and black Hole masses of the identified LINERs}\\
\label{liner}\\
\hline
SDSS name&$lg {L}_{NLR}$&${\sigma}_{*}$&$lg {M}_{BH}$\\
(1)&(2)&(3)&(4)\\
\hline
\endfirsthead
\hline
SDSS name&$lg {L}_{NLR}$&${\sigma}_{*}$&$lg {M}_{BH}$\\
(1)&(2)&(3)&(4)\\
\hline
\endhead
\hline
J085004.65+403607.7& 42.21&412$\pm$22& 9.4\\
J093430.68+030545.3& 42.53&393$\pm$16& 9.3\\
J111421.76+582319.8& 42.11&          &     \\
J111908.94+090022.8& 41.96&315$\pm$29& 8.9\\
J113255.96+051539.6& 41.73&231$\pm$ 7& 8.4\\
J113518.79+125311.1& 42.10&          &     \\
J132419.67+041907.0& 42.29&375$\pm$38& 9.2\\
J150324.77+475829.6& 42.68&          &     \\
J151838.90+404500.2& 41.55&241$\pm$ 7& 8.5\\
J153253.78+302059.3& 43.57&          &     \\
J160239.61+264606.0& 43.15&429$\pm$87& 9.5\\
J172010.03+263732.0& 42.26&364$\pm$19& 9.2\\
\hline
\end{longtable}
%l\tablenotetext{
\tablecomments{0.53\textwidth}{
 (1): SDSS name --- Jhhmmss.s+ddmmss.s\\
 (2): $lg L_{NLR}$ --- log narrow emission line region luminosity in Unit: $erg\,s^{-1}$\\
 (3): ${\sigma}_{*}$ --- velocity dispersion and velocity dispersion error of the host galaxy in Unit: $km\,s^{-1}$\\
 (4): $lg M_{BH}$ --- log Black Hole mass in Unit: Solar mass\\
 }
%}

\begin{longtable}{cccc}
\caption{Comparison with previous RGB and ROXA identifications}\\
\label{tab_comp}\\
\hline
SDSS name&our ID&previous ID&ref\\
(1)&(2)&(3)&(4)\\
\hline
\endfirsthead
\hline
SDSS name&our ID&previous ID&ref\\
(1)&(2)&(3)&(4)\\
\hline
\endhead
\hline
 J083353.88+422401.8 &BL Lac&FSRQ&2\\
 J111908.94+090022.8 &LINERs&BL Lac&2\\
 J113518.79+125311.1 & LINERs&Radio Galaxy&2\\
 J115409.28+023815.1 &FSRQ&FSRQ/BL Lac&2\\
 J120303.50+603119.1 &BL Lac&LINERs&3,2\\
 J124834.30+512807.8 &BL Lac&Star&1,2,4\\
 J132419.67+041907.0 &LINERs&SSRQ&2\\
 J141149.43+524900.1 &Galaxy&BL Lac&2\\
 J142730.27+540923.7 &Galaxy&BL Lac&1\\
 J150117.97+545518.2 &Seyfert 2&SSRQ&2\\
 J153253.78+302059.3 &LINERs&NELG&2\\
\hline
\end{longtable}
%\tablecomments{
\tablecomments{0.53\textwidth}{
 (1): SDSS name --- Jhhmmss.s+ddmmss.s\\
 (2): our ID --- our SDSS spectral class\\
 (3): previous ID --- previous spectral class\\
 (4): ref --- reference about source spectral identification\\
              1:Laurent-Muehleisen et~al.~\cite{LM98} \\
              2:Turriziani et~al.~\cite{Tur07}\\
              3:Carrillo et~al.~\cite{Car99}\\
              4:Collinge et~al.~\cite{Collinge05}\\
              }
%}

\begin{longtable}{ccccccl}
\caption{BL Lac with a featureless continuum}\\
\label{bllac0}\\
\hline
SDSS name&$u$&$g$&$r$&$i$&$z$&ref\\
(1)&(2)&(3)&(4)&(5)&(6)&(7)\\
\hline
\endfirsthead
\hline
SDSS name&$u$&$g$&$r$&$i$&$z$&ref\\
(1)&(2)&(3)&(4)&(5)&(6)&(7)\\
\hline
\endhead
\hline
 J085409.88+440830.2& 17.33& 17.02& 16.81& 16.67& 16.55&1,2\\
 J092915.43+501336.1& 17.26& 16.86& 16.49& 16.22& 16.01&2\\
 J100110.20+291137.6& 19.02& 18.52& 18.08& 17.75& 17.41&\\
 J103744.30+571155.6& 17.07& 16.68& 16.38& 16.13& 15.91&1,2\\
 J110124.72+410847.4& 19.35& 19.06& 18.80& 18.58& 18.37&2\\
 J110748.07+150210.5& 18.53& 18.27& 18.05& 17.88& 17.71&\\
 J112453.82+493409.7& 18.96& 18.70& 18.50& 18.37& 18.15&2\\
 J115124.67+585917.7& 16.40& 16.08& 15.89& 15.64& 15.58&2\\
 J120922.78+411941.3& 18.38& 17.99& 17.61& 17.34& 17.10&2\\
 J124700.72+442318.7& 19.02& 18.84& 18.50& 18.37& 18.12&2\\
 J130145.65+405624.6& 18.55& 18.28& 18.04& 17.91& 17.76&2\\
 J135120.84+111453.0& 18.99& 18.72& 18.50& 18.21& 18.04&2\\
 J141536.80+483030.4& 19.61& 19.18& 18.74& 18.36& 18.11&1\\
 J142607.71+340426.3& 17.95& 17.59& 17.33& 17.09& 16.90&\\
 J144052.93+061016.1& 17.77& 17.38& 17.08& 16.83& 16.63&\\
 J144800.58+360831.0& 17.32& 16.96& 16.75& 16.50& 16.41&1,2\\
 J145427.13+512433.7& 18.58& 18.17& 17.77& 17.46& 17.20&1,2\\
 J150947.97+555617.2& 18.44& 18.03& 17.72& 17.47& 17.21&2\\
 J153324.26+341640.3& 18.29& 18.04& 17.86& 17.65& 17.49&1,2\\
 J160218.06+305109.4& 18.49& 18.13& 17.91& 17.77& 17.59&1,2\\
 J161830.59+062211.5& 19.25& 18.84& 18.46& 18.16& 17.97&\\
 J165249.92+402310.1& 18.90& 18.57& 18.31& 18.12& 18.04&1,2\\
\hline
\end{longtable}
%l\tablenotetext{
\tablecomments{0.60\textwidth}{
(1): SDSS name --- Jhhmmss.s+ddmmss.s\\
% (2): $u$ --- Galactic extinction corrected psf magnitude of SDSS u band\\
% (3): $g$ --- Galactic extinction corrected psf magnitude of SDSS g band\\
% (4): $r$ --- Galactic extinction corrected psf magnitude of SDSS r band\\
% (5): $i$ --- Galactic extinction corrected psf magnitude of SDSS i band\\
% (6): $z$ --- Galactic extinction corrected psf magnitude of SDSS z band\\
(2)-(6): Galactic extinction corrected psf magnitude of SDSS\\
(7): ref --- reference about source spectral identification\\
              No reference suggests that it is firstly identified by our work.\\
              1:Laurent-Muehleisen et~al.~\cite{LM98} \\
              2:Turriziani et~al.~\cite{Tur07}\\
              }
%}
\end{center}
\appendix
\section{SDSS starlight spectral modeling}

The spectra are first corrected for Galactic extinction using the
extinction map of Schlegel et~al.~(\cite{Schlegel98}) and the
reddening curve of Fitzpatrick~(\cite{fitz99}), and transformed into
the rest frame using the redshift provided by the SDSS pipeline.
Then, host-galaxy starlight and AGN continuum, as well as the
optical Fe II emission complex are modeled as
\[
S(\lambda)=A_{host}(E_{B-V}^{host},\lambda)~A(\lambda)
+A_{nucleus}(E_{B-V}^{nucleus},\lambda)~ [bB(\lambda)+ c_{\rm b}
C_{\rm b}(\lambda) + c_{\rm n} C_{\rm n}(\lambda) ]
\]
where $S(\lambda)$ is the observed spectrum.
$A(\lambda)=\sum_{i=1}^6 a_{i}~IC_{i}(\lambda,\sigma_{*})$
represents the starlight component modeled by our six synthesized
galaxy templates, which was built up from the spectral template
library of Simple Stellar Populations (SSPs) of Bruzual \&
Charlot~(\cite{bc03}) using our new method based on the Ensemble
Learning Independent Component Analysis (EL-ICA) algorithm. The
details of the galaxy templates and their applications are presented
in Lu et~al.~(\cite{lu06}). $A(\lambda)$ is broadened by convolving
it with a Gaussian of width $\sigma_{*}$ to match the stellar
velocity dispersion of the host galaxy. The un-reddened nuclear
continuum is assumed to be $B(\lambda)=\lambda^{-1.7}$ as given in
Francis ~(\cite{fr96}). We modeled the optical Fe II emission, both
broad and narrow, using the spectral data of the Fe II multiplets
for I\,Zw\,I in the $\lambda\lambda$\,3535--7530\AA\ range provided
by V\'{e}ron-Cetty et~al.~(\cite{veron04})(Table\,A1,A2). We assume
that the  broad Fe II lines ($C_{\rm b}$ in Eq.(A.1)) have the same
profiles as the broad \hb line, and the narrow Fe II lines ($C_{\rm
n}$), both permitted and forbidden, have the same profiles as the
that of the narrow \hb component, or of [O III]$\lambda5007$ if \hb
is weak. $A_{host}(E_{B-V}^{host},\lambda)$ and
$A_{nucleus}(E_{B-V}^{nucleus},\lambda)$ are the color excesses due
to possible extinction of the host galaxy and the nuclear region,
respectively, assuming the extinction curve for the Small Magellanic
Cloud of Pei~(\cite{pei92}). The fitting is performed by minimizing
the $\chi^{2}$ with $E_{B-V}^{host}$, $E_{B-V}^{nucleus}$, $a_{i}$,
$\sigma_{*}$, $b$, $c_{\rm b}$ and $c_{\rm n}$ being free
parameters. To account for possible errors of the redshifts provided
by the SDSS pipeline, in practice, we iterate the procedure with
redshifts with a step of 5 \kmps near the SDSS redshift  and use the
best-fit result.

The emission line spectra are fitted in the following way. As the
narrow Balmer lines and the [N II]$\lambda\lambda$6548, 6583 doublet
have similar profiles to the [S II]$\lambda\lambda$6716, 6731
doublet lines (Filippenko et~al.~\cite{Filippenko88}; Ho et
al.~\cite{b11b}; Zhou et~al.~\cite{zhou06}), we use the [S
II]$\lambda\lambda$6716, 6731 doublet lines as a template to fit the
narrow lines. If [S II] is weak, [O III]$\lambda$5007 is used as the
template. Each of the [S II] doublet lines is assumed to have the
same profile and redshift, and is fitted with as many Gaussians as
is statistically justified; generally 1--2 Gaussians are needed.
Likewise, the [O III]$\lambda\lambda$4959, 5007 doublet are fitted
in a similar way, with the flux ratio of [O III]$\lambda$5007/[O
III]$\lambda$4959 fixed to the theoretical value of 3. When a good
model of the narrow-line template is achieved, we scale it to fit
the narrow Balmer lines and the [N II]$\lambda\lambda$6548, 6583
doublet lines. The flux ratio of the [N II] doublet
$\lambda$6583/$\lambda$6548 is fixed to the theoretical value of
2.96. For possible broad \hae\, and \hbe\, lines, we use multiple
Gaussians to fit them, as many as is statistically justified. If a
broad \hbe\, line is too weak to get a reliable fit, we then re-fit
it assuming it has the same profile and redshift as a broad \hae. If
a broad-line emission is detected at the $\geq 5 \sigma$ confidence
level, we regard it as genuine.


\label{lastpage}


\begin{thebibliography}{99}
%% you can type \apj for ApJ, \aap for A&A, \apss for Ap&SS, etc. Please consult
%% the macro raa.cls. You can also find them in aasguide.tex (AASTeX for ApJ, AJ, PASP)
%% Please follow the formats of RAA's references list as demonstrated below:
\bibitem[2009]{abdo}Abdo, A. A., Ackermann, M., Ajello, M., et~al., 2009, ApJ, 699, 976
\bibitem[1981]{bpt}Baldwin, J. A., Phillips, M. M., Terlevich, R., 1981, PASP, 93, 5
\bibitem[1995]{Brinkmann95}Brinkmann, W., Siebert, J., Reich, W. et~al., 1995, A\&AS, 109, 147
\bibitem[1997]{Brinkmann97}Brinkmann, W., Siebert, J., Feigelson, E. D. et~al., 1997, A\&A, 323, 739
\bibitem[2003]{bc03}Bruzual, G., Charlot, S., 2003, MNRAS, 344, 1000
\bibitem[1999]{Caccianiga99}Caccianiga, A., Maccacaro, T., Wolter, A. et~al., 1999, ApJ, 513, 51
\bibitem[2000]{Caccianiga2000}Caccianiga, A., Maccacaro, T., Wolter, A. et~al., 2000, A\&AS, 144, 247
\bibitem[1999]{Car99}Carrillo R., Masegosa, J., et~al., 1999, Rev.Mex.AA, 35, 187
\bibitem[1993]{cel93}Celotti, A., Fabian, A. C.,  1993, MNRAS, 264, 228
\bibitem[1997]{cel97}Celotti, A., Padovani, P.,  \& Ghisellini, G. 1997, MNRAS, 286, 415
\bibitem[1977]{cos77}Costero, R. \& Osterbrock, D.E., 1977, ApJ, 211, 675
\bibitem[2005]{Collinge05}Collinge, M. J., Strauss, M. A., Hall, P. B. et~al., 2005, AJ, 129, 2542
\bibitem[1999]{craw}Crawford, C. S., Allen, S. W., Ebeling, H., Edge, A. C., Fabian, A. C, 199, MNRAS, 306, 857
\bibitem[2005]{dong05} Dong, X.B., Zhou, H.Y., Wang, T.G. et~al., 2005, ApJ, 620, 629
\bibitem[1987]{ds87} Dressler, A., Shectman, S. A., 1987, AJ, 94, 899
\bibitem[2003]{edge2003}Edge, A. C., Ebeling, H., Bremer, M. et~al., 2003, MNRAS, 339, 913
\bibitem[2000]{fal00} Falcke, H., et al., 2000, ApJ, 542, 197
\bibitem[1988]{Filippenko88}Filippenko, A. V., Sargent, W.L.W., et~al., 1988, ApJ, 324, 134
\bibitem[1999]{fitz99}Fitzpatrick, E. L., 1999, PASP, 111, 63
\bibitem[1998]{Fos98}Fossati, G., Maraschi, L., Celotti, A. et~al., 1998, MNRAS, 299, 433
\bibitem[1991]{fr91}Francis, P. J., Hewett, P. C., Foltz, C. B. et~al., 1991, ApJ, 373, 465
\bibitem[1996]{fr96} Francis, P. J., 1996, Publ.Astron.Soc.Australia, 13, 212
\bibitem[1981]{gas81}Gaskell, C. M., Sheilds, G. A., Wampler, E. J., 1981, ApJ, 249, 443
\bibitem[1998]{ghi98} Ghisellini, G., Celotti, A., Fossati, G. et~al., 1998, MNRAS, 301, 451
\bibitem[1999]{Gio99a} Giommi, P., Menna, M.~T.,  \& Padovani, P, 1999, MNRAS, 310, 465
\bibitem[2005]{Gio05}Giommi, P., Piranomonte, S., Perri, M. et~al., 2005, A\&A, 434, 385
\bibitem[2005]{gh05}Greene, J. E., Ho, L. C., 2005, ApJ, 630, 122
\bibitem[2007]{gh07}Greene, J. E., Ho, L. C.,  2007a, ApJ, 670, 92
\bibitem[2007]{greene07}Greene, J. E., Ho, L. C.,  2007b, ApJ, 667, 113
\bibitem[1991]{Gre91}Gregory P. C., Condon J. J., 1991, ApJS, 75. 1011
\bibitem[1996]{Gre96} Gregory P. C., Scott W. K., Douglas K., 1996, ApJS, 103, 427
\bibitem[1980]{h80} Heckman T. M., 1980, A\&A, 87, 152
\bibitem[1979]{hin79} Hine, R.G. \& Longair, M.S., 1979, MNRAS, 188, 111
\bibitem[1995]{ho95} Ho, L.C., Filippenko, A.V., \& Sargent,W. L.W., 1995, ApJS, 98, 477
\bibitem[1997]{b11b} Ho, L. C., et al., 1997, ApJs, 112, 315
\bibitem[2008]{ho08} Ho, L.C.,\, 2008, ARA\&A, 46, 475
\bibitem[2005]{Jester2005} Jester, Sebastian, 2005, AJ, 130, 873J
\bibitem[2003]{k03}Kauffmann, G., Heckman, T. M., Tremonti, C. et~al., 2003, MNRAS, 346, 1055
\bibitem[1989]{kel89}Kellermann, K. I., Sramek, R., Schmidt, M. et~al., 1989, AJ, 98, 1195
\bibitem[2006]{k06}Kewley, L. J., Groves, B., Kauffmann, G. et~al., 2006, MNRAS, 372, 961
\bibitem[2001]{k01}Kewley, L. J., Dopita, M. A., Sutherland, R. S. et~al., 2001, ApJ, 556, 121
\bibitem[2006]{kom06}Komossa, S., Voges, W., Xu, D. et~al.,2006, AJ, 132, 531
\bibitem[2007]{koes1}Koester, B. P., McKay, T. A., Annis, J. et~al., 2007, ApJ, 660, 239
\bibitem[1978]{kos78} Koski, A.T., 1978, ApJ, 223, 56
\bibitem[1994]{lai94} Laing, R.A., et al.\ 1994, 'The physics of Active Galaxies', Bicknell G.V., Dopita, M.A., Quinn P.J. (eds.) ASPC, 54, 201
\bibitem[2001]{L01} Landt, H., Padovani, P., Perlman, E.~S. et~al., 2001, MNRAS, 323, 757
\bibitem[2002]{L02} Landt, H., Padovani, P., Giommi, P., 2002, MNRAS, 336, 945
\bibitem[1997]{LM97} Laurent-Muehleisen, S. A., Kollgaard, R. I., Ryan, P. J. et~al., 1997, A\&AS, 122, 235
\bibitem[1998]{LM98}Laurent-Muehleisen, S.~A., Kollgaard, R.~I., Ciardullo, R. et~al., 1998, ApJS, 118, 127
\bibitem[1999]{LM99} Laurent-Muehleisen, S.~A., Kollgaard, R., Feigelson, E.~D. et~al., 1999, ApJ, 525, 127
\bibitem[1985]{led85}Ledden, J. E., Odell, S. L., 1985, ApJ, 298, 630
\bibitem[2003]{lew03} Lewis, L., Eracleous, M., Sambruna, R.M., 2003, ApJ, 593, 115
\bibitem[2006]{lu06}Lu, H.L, Zhou, H.Y., et al., 2006, AJ, 131, 790
\bibitem[2008]{mar08} Maraschi, L., et al., 2008, MNRAS, 391, 1981
\bibitem[1996]{Mar96}Marcha, M. J. M., Browne, I. W. A., Impey, C. D et al., 1996, MNRAS, 281, 425
\bibitem[2004]{Mcl2004}McLure, R. J., Dunlop, J. S. 2004, MNRAS, 352, 1390
\bibitem[2008]{McG08}McGill, K. L., Woo, J.H., Treu, T. et~al., 2008, ApJ, 673, 703
\bibitem[2000]{nag00} Nagar, N.M., et al., 2000, ApJ, 542, 186
\bibitem[2005]{nag05} Nagar, N.M., Falcke, H., \& Wilson, A.S., 2005, A\&A, 435, 521
\bibitem[1976]{ost76} Osterbrock D.E., 1976, PASP, 88, 589
\bibitem[1985]{op85}Osterbrock, D. E., Pogge, R. W., 1985, ApJ, 297, 166
\bibitem[2003]{P03}Padovani, P., Perlman, E.~S., Landt, H. et~al., 2003, ApJ, 588, 128
\bibitem[1992]{pei92}Pei, Y. C., 1992, ApJ, 395, 130
\bibitem[1998]{Per98}Perlman, E.~S., Padovani, P., Giommi, P. et~al., 1998, AJ, 115, 1253
\bibitem[1997]{peterson} Peterson, B. M., 1997, An Introduction to Active Galactic Nuclei(Cambridge: Cambridge Univ.Press)
\bibitem[2007]{Pir07}Piranomonte S., Perri M., Giommi P. et al, 2007, A\&A, 470, 787
\bibitem[2006]{r06}Richards, G. T., Lacy, M., Storrie-Lombardi, L. J., et~al., 2006, ApJS, 166, 470
\bibitem[1989]{sad89} Sadler, E.M., Jenkins, C.R., Kotanyi, C.G., 1989, MNRAS, 240, 591
\bibitem[1998]{Schlegel98}Schlegel, D. J., Finkbeiner, D. P., Davis, M., 1998, ApJ, 500, 525
\bibitem[1991]{sto91}Stocke, J.T., Morris, S.L., Gioia, I. M., et al, 1991, ApJS, 76, 813
\bibitem[2002]{st02}Stoughton, C., Lupton, R. H., Bernardi, M. et~al., 2002, AJ, 123, 485
\bibitem[1998]{tad98} Tadhunter, C.N., et al., 1998, MNRAS, 298, 1035
\bibitem[2002]{trem02}Tremaine, S., Gebhardt, K., Bender, R. et~al., 2002, ApJ, 574, 740
\bibitem[2007]{Tur07}Turriziani, S., Cavazzuti, E., Giommi, P., 2007, A\&A, 472, 699
\bibitem[1995]{urry95} Urry, C.M. \& Padovani, P., 1995, PASP, 107, 803
\bibitem[2004]{Van04} Vanden~Berk, D.~E., et~al., 2004, ApJ, 601, 692
\bibitem[1987]{vo87} Veilleux, S. \& Osterbrock, D. E.,  1987, ApJS, 63, 295
\bibitem[2004]{veron04}V\'{e}ron-Cetty, M.-P., Joly, M., V\'{e}ron, P., 2004, A\&A, 417, 515
\bibitem[2006]{vp06}Vestergaard, M., Peterson, B. M., 2006, ApJ, 641, 689
\bibitem[1999]{vog} Voges, W., Aschenbach, B., Boller, Th. et~al, 1999, A\&A, 349, 389
\bibitem[2006]{wha06} Whalen, D. J., Laurent-Muehleisen, S. A. et~al, 2006, AJ, 131, 1948
\bibitem[2008]{yuan08} Yuan, W., Zhou, H.Y., \& Komossa, S. et al., 2008, ApJ, 685, 801
\bibitem[2006]{zhou06} Zhou, H.Y., Wang, T.G., Yuan, W.M. et~al., 2006, ApJS, 166, 128
\bibitem[1995]{zb95} Zirbel, E.L., Baum, S.A., 1995, ApJ, 448, 521
\end{thebibliography}
\end{document}